\documentclass[review,number,sort&compress]{elsarticle}
\usepackage{lineno}
\usepackage{graphicx}% Include figure files
\usepackage{dcolumn}% Align table columns on decimal point
\usepackage{bm}% bold math
\usepackage{amsmath}
\usepackage{multirow}

\newcommand{\mean}[1]{\mbox{$\langle{#1}\rangle$}}

\usepackage[colorlinks=true]{hyperref}

\begin{document}
\date{\today}
%\title{Longitudinal space charge amplifier at Femilab Accelerator Science and Technology facility (FAST)}
\title{Simulation of a cascaded longitudinal space charge amplifier for coherent radiation generation}

\author[niu,apc]{A. Halavanau\corref{cor1}}
\author[niu,apc]{P. Piot}
\address[niu]{Department of Physics and Northern Illinois Center for Accelerator \& Detector Development, \\
Northern Illinois University, DeKalb, IL  60115, USA}
\address[apc]{Accelerator Physics Center, Fermi National Accelerator Laboratory, Batavia, IL  60510, USA}
\cortext[cor1]{corresponding author, Tel: +1 815 753 6415, E-mail addresses: aliaksei.halavanau@gmail.com, Z1720441@students.niu.edu (A. Halavanau).}
\date{today}

\begin{abstract}

Longitudinal space charge (LSC) effects are generally considered as harmful in free-electron lasers as they can seed unfavorable energy modulations that can result in density modulations with associated emittance dilution. This ``micro-bunching instabilities'' is naturally broadband and could possibly support the generation of coherent radiation over a broad region of the spectrum. Therefore there has been an increasing interest in devising accelerator beam lines capable of controlling LSC induced density modulations. 
In the present paper we refine these previous investigations by combining a grid-less space charge algorithm with the popular particle-tracking program {\sc elegant}. This high-fidelity model of the space charge is used to benchmark conventional LSC models. We finally employ the developed model to investigate the performance of a cascaded LSC amplifier using beam parameters comparable to the ones achievable at Fermilab Accelerator Science \& Technology (FAST) facility currently under commissioning at Fermilab. 

\end{abstract}

%
%\pacs{ 29.20.Ej, 29.27.-a, 41.85.-p,  41.75.Fr}
\begin{keyword} Linear accelerator \sep Electron beams \sep Space charge \sep Micro-bunching instabilities 
\sep $N$-body tree algorithm
\end{keyword}
\maketitle

%\linenumbers

\section{Introduction}
Longitudinal-space-charge-driven micro-bunching instabilities arising in bunch compressors were predicted and observed over the last decade~\cite{Loos,Piot:2000,Borland:2008}. It was recently proposed to employ 
such micro-bunching instability mechanism to form attosecond structures on the bunch 
current distribution for the subsequent generation of coherent radiation pulses~\cite{Dohlus:2011}. 

\begin{figure}[hhh!!!!!!]
\includegraphics[width=0.98\linewidth]{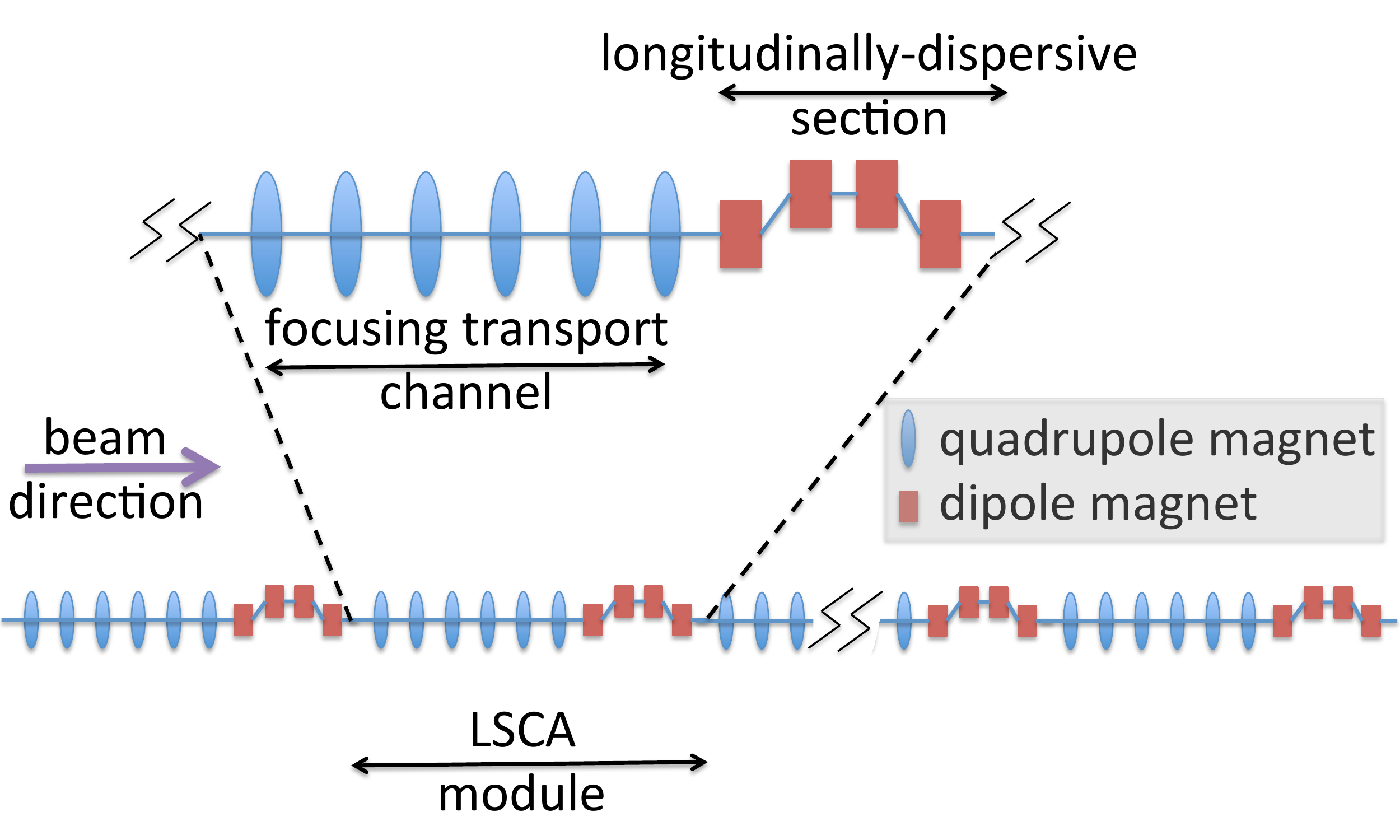}
\caption{\label{lsca-layout}  Overview of a cascaded longitudinal-space-charge amplifier (LSCA) composed of several LSCA modules. Each LSCA module incorporate a focusing channel and a longitudinally dispersive section. The (red) rectangles and (blue) ellipses respectively represent dipole and quadrupole magnets. }
\end{figure}

A possible  beam line configuration capable of  enabling the micro-bunching instability is relatively simple. It  essentially consists of focusing section (e.g. FODO cells) where energy modulations due to the LSC  impedance accumulate, followed by a longitudinally-dispersive section. The latter  section, by introducing an energy dependent path length, converts the incoming energy modulation into a density modulation. Such an elementary cell is often referred to as a LSC amplifier (LSCA). 
Most of the beamlines studied so far consider a longitudinally-dispersive section arranged as a bunch compression chicane [or bunch compressor (BC)]; see Fig.~\ref{lsca-layout}. Several of these LSCA modules are concatenated so to result in a large final density modulation. We further assume the compression process in the chicane is linear [the incoming longitudinal phase space (LPS) does not have any nonlinear correlations]. Such a modulated beam, when  participating in a radiation-generation process, can produce coherent radiation at wavelengths comparable to the spectral range of the final density modulations. 

The purpose of this paper is two-fold. The paper first introduces a fully three dimensional (3D) 
multi-scale space-charge algorithm adapted from Astrophysics~\cite{Barnes:1986}. The algorithm is used to discuss 
some limitations of the one-dimensional LSC impedance model commonly employed in LSCA investigations.
Using the latter benchmarked algorithm, we then investigate a possible LSCA beamline configuration similar to the one studied in~\cite{Dohlus:2011}. Finally, we estimate the generation of undulator radiation seeded by the LCSA. In contrast to Ref. ~\cite{Dohlus:2011} our study consider the case of a $\sim 500$~A 300-MeV electron beam produced in a conventional superconducting linac.  

\section{Mechanism for longitudinal space charge amplifiers}

Charged-particle beams are subject to self interaction via velocity and radiation fields. In absence of radiation 
processes (i.e. acceleration), the effect of velocity fields (i.e. space charge) dominates and its regime varies with 
the bunch density. Under a simple 1D approximation, a comparison of the  Debye length $\lambda_D$ to the root-mean-squared (rms) 
transverse beam size $\sigma_{\perp}$ and mean inter-particle distance $\Lambda_p\simeq n_e^{-1/3}$ (where $n_e$ is the electronic density)  provides a criterion to assess the importance of space charge effects on the beam dynamics. 
When $\sigma_{\perp}< \lambda_D$ space charge effects are significant and often computed using the mean-field  approximation (i.e. 
the space charge force is derived from the electrostatic potential associated to the particle distribution) commonly implemented 
in particle-in-cell (PIC) algorithms. However, when $\lambda_D\sim {\cal{O}}( \Lambda_p)$, particle-to-particle ``binary'' 
interactions play an important role and are needed to be accounted for~\cite{reiser}.

As the beam is accelerated the transverse and longitudinal space-charge forces reduce respectively as ${\cal O} (1/\gamma^2)$ and ${\cal O} (1/\gamma^3)$ where $\gamma$ is the Lorentz factor. At the macroscopic level, e.g. for spatial scale comparable to the bunch sizes, the space charge can be accurately described by a mean field approach~\cite{Burov:2009}. However, 
in high-brightness beams $-$ beams with low fractional momentum spread $-$ the weakened longitudinal-space charge (LSC) force can 
still influence the beam dynamics at a microscopic level $-$ i.e. for spatial scales smaller than the bunch sizes $-$ and small density modulations (e.g. due to noise or imperfections) can result in LCS-driven energy modulations. In this latter regime, the LSC is generally treated with a one-dimensional (1D) model.

To illustrate the main results of the 1-D model, we  consider a simple beam line consisting of a drift with length $L_d$ (where the beam is transversely contained) followed by a chicane with longitudinal dispersion $R_{56}$. It is customary to characterize the strength of the micro-bunching instability by associating the density gain defined as 
\begin{eqnarray}
G(k) &=&  \frac{b_i(k)}{b_f(k)}, 
\end{eqnarray}
where $k \equiv \frac{2\pi}{\lambda}$ and $\lambda$ is the observation wavelength and $b_{i,f}$ are respectively the initial and final bunching factors
defined as 
\begin{eqnarray}
 b(\omega)=\frac{1}{N}\bigg| \sum_{n}\exp(-i\omega t_n)\bigg|,
\end{eqnarray}
where $t_n$ is the temporal coordinate of the $n$-th macroparticle, $N$ is the total number of particles and $\omega \equiv k c$. In the latter equation we assume the beam's longitudinal density to follow the Klimontovich distribution $\rho(t)=\frac{1}{N}\sum_{j=1}^N \delta (t-t_j)$.

The gain for this simple beam line can be shown to be proportional to the impedance $Z(k,r)$ \cite{Saldin2002516} following 
\begin{eqnarray}
\label{gaineq}
G=C k |R_{56}|\frac{I}{\gamma I_A}\frac{4\pi L_d |Z(k,r)|}{Z_0}e^{-\frac{1}{2}C^2 k^2 R_{56}^2 \sigma_\delta^2}, 
\end{eqnarray}
where $I_A=17$~kA is the Alfv\`en current,  $\sigma_\delta$ is the rms fractional energy spread, $C\equiv \langle z\delta\rangle/\sigma_z$ is the chirp,  
and $Z_0\equiv 120\pi$ is the free-space impedance. 

The exponential term in Eq.~\ref{gaineq} induces a high-frequency cut-off of the modulation 
\begin{eqnarray}\label{eqn:cutoff}
R_{56}\approx -\frac{c}{\omega \sigma_\delta}.
\end{eqnarray}
Note, that after traveling through a BC, the modulation wavelength will be shortened by a compression factor $\kappa\equiv (1+R_{56}C)$. Although the impedance $Z(k,r)$ is partially determined by the properties of the wakefields inside the BC~\cite{Saldin2002516}, the LSC has much stronger effect in amplifying density modulations~\cite{Dohlus:2011,Dohlus:2013}.

For a transversely Gaussian cylindrically-symmetric beam the LSC impedance is given by \cite{Venturini:2008}
\begin{equation}
\label{imp_ven}
 Z(k)=-i\frac{Z_0}{\pi \gamma \sigma_\perp}\frac{\xi_{\sigma_\perp}}{4}e^{\xi_{\sigma_\perp}^2/2}\mbox{Ei}(-\frac{\xi_{\sigma_\perp}^2}{2}),
\end{equation}
where $Z_0=120\pi$  is the free-space impedance, $\mbox{Ei}(x)\equiv -\int_{-x}^{\infty}dt e^{-t}/t$,  $\sigma_\perp$ is 
the rms beam size and $\xi_{\sigma_\perp}\equiv k \sigma_\perp / \gamma$.  Similar expression for a transversely uniform beam is provided in \cite{Huang:2004}.

The maximum of the Eq. \ref{imp_ven} is achieved at $\xi_{\sigma_\perp}\approx 1$, therefore the optimal wavelength of the density modulation will be located around 
\begin{eqnarray}\label{eqn:lambdaopt}
\lambda_{opt}=2\pi \sigma_\perp/\gamma. 
\end{eqnarray}

\section{Simulation procedure and benchmarking}
The nature of space charge forces lies in particle-to-particle Coulomb interaction. Direct summation of the forces yields to ${\cal O}(N^{2})$ growth where $N$ is the number of macroparticles, which makes it impossible to compute at large $N$.
 Several approximation techniques can be used: mean-field on a grid approximation \cite{ASTRAmanual}, one-dimensional space charge impedance \cite{Venturini:2008},  analytical sub-beams or ensembles model \cite{Krasilnikov200669}, rigid-slice approximation \cite{Burov:2009}. All of those methods reduce the problem's complexity via some approximations which ultimately limits the maximum attainable spatial resolution. Most recent attempt used a three-dimensional-grid space charge algorithm based on a periodic boundary~\cite{Dohlus:2011}.

From another point of view, space charge problem is very similar to the 
well-known $N$-body problem in celestial mechanics.  One of the most effective algorithms for 
the gravitational $N$-body problem  is the so called ``tree'' or Barnes-Hut (BH) algorithm~\cite{Barnes:1986}, which scales as ${\cal O}(N \log N)$. 
In this  paper we present the results obtained  using  a modified version of the code available at~\cite{treemanual}. 
Such approach was successfully employed to simulate early-stage beam dynamics in photocathodes~\cite{Maxson:1538515} and laser ion cooling~\cite{vanderGeer}.  

In brief, the BH algorithm initially surrounds the bunch distribution in a cubic cell called a root cell. 
The root cell is divided into 8 sub-cells recursively, until it reaches the point where a single sub-cell contains just one particle. 
Then forces only between nearby cells are calculated directly, and the cells far away from each other are treated as two 
large macroparticles with the total charge placed in the cell's center of mass. The process of calculating net forces starts 
from the root cell and recursively parses the cell hierarchy until it reaches the size of the smallest cell that is predefined as a 
precision parameter. Thus, the algorithm is significantly faster than a direct summation method. The BH method does not preserve full 
Hamiltonian, yet for relatively small precision parameter the difference between direct summation is comparably small~\cite{Barnes:1986}. 
It should be pointed out in the direct summation part (for neighboring cells) the BH algorithm also implement a local smoothing of the 
potential to avoid singularities~\cite{Barnes:1986}. 

Another more efficient Fast Multipole Method (FMM) algorithm has been recently developed~\cite{Bela:alg, Bela2:alg} and will be 
eventually used in further refinement of our work. Though FMM algorithms are more sophisticated and precise, still the BH method 
is comparably accurate and faster than FMM and it can be embedded in time-stepping integrators \cite{Winkel2012880}.

For the studies detailed in this paper we used the BH algorithm as an external script within the {\sc elegant} simulations. At a 
user-specified axial locations along  the accelerator beam line, space charge kicks were applied. Our approach is to follow 
the quasi-static approximation~\cite{qj}. The distribution at the defined axial location is recorded and a Lorentz transformation 
to the bunch's rest frame is performed. The BH algorithm is then utilized to obtain the 3D electrostatic field $\mathbf{E'}$ in the bunch's rest frame. We should point out that the BH algorithm returns the field directly evaluated at each macro-paticle locations (so that there is not need for interpolation as in a grid-based algorithm). The electrostatic field is subsequently transformed in the laboratory frame via a Lorentz boost.  The resulting  electromagnetic fields $(\mathbf{E},\mathbf{B})$ in the laboratory frame are used to apply the corresponding Lorentz force $\mathbf{F}=q [\mathbf{E} + c\pmb{\beta} \times \mathbf{B}]$ on each of the macroparticles [$q$ and $\pmb{\beta}$ are respectively the charge and reduced velocity ($\beta \equiv \sqrt{1-\gamma^{-2}}$) of the considered macroparticle]. 

The distribution is finally passed back to {\sc elegant} and tracked up in the given optical lattice to the next space charge kick where the above process repeats. We henceforth refer to the combination of the BH algorithm  with {\sc elegant} as {\sc ``elegant-bh''}.

Our implementation relies on an impulse approximation so that only the momentum, i.e. not the position, of the macroparticles is altered by the space charge kick. We assumed there is no magnetic field in the rest frame. Although this assumption is not strictly valid, it was shown to hold for beams with low energy spread typically generated from photoinjectors~\cite{Candel:2008}. 

In order to gain confidence in the implemented space charge calculation procedure, several validation tests were conducted; see ~\ref{app:benchmarks}. In this section we only focus on the benchmarking of Eq.~\ref{imp_ven} with {\sc elegant-bh}.  We considered initial bunch distributions with modulated current profiles of the form 
\begin{eqnarray}
f(\mathbf x)&=&T(x,y)L(z)\left[1+m\cos{kz}\right],
\end{eqnarray}
 where $\mathbf x \equiv (x,y,z)$, $m$ and $k$ are respectively the modulation amplitude and spatial wavenumber, and $L(z)$ and $T(x,y)$ are respectively the nominal longitudinal 
and transverse beam distributions. 

The modulation along the axial $z$ direction leads to an energy modulation due to the LSC impedance and eventually produces further current modulation depending on the longitudinal dispersion of the beamline. From the definition of the impedance, and given the Fourier-transformed longitudinal 
electric field $\widetilde{E}_z(k)$ and current distribution $\widetilde{I}(k)$, the longitudinal impedance can be recovered as 
\begin{eqnarray}\label{eq:impedanceretr}
Z(k)= - \frac{\widetilde{E}_z(k)}{ \widetilde{I}(k)}. 
\end{eqnarray} 

\begin{figure}[hhhhh!!!!!]
\begin{center}
\includegraphics[width=1.0\linewidth]{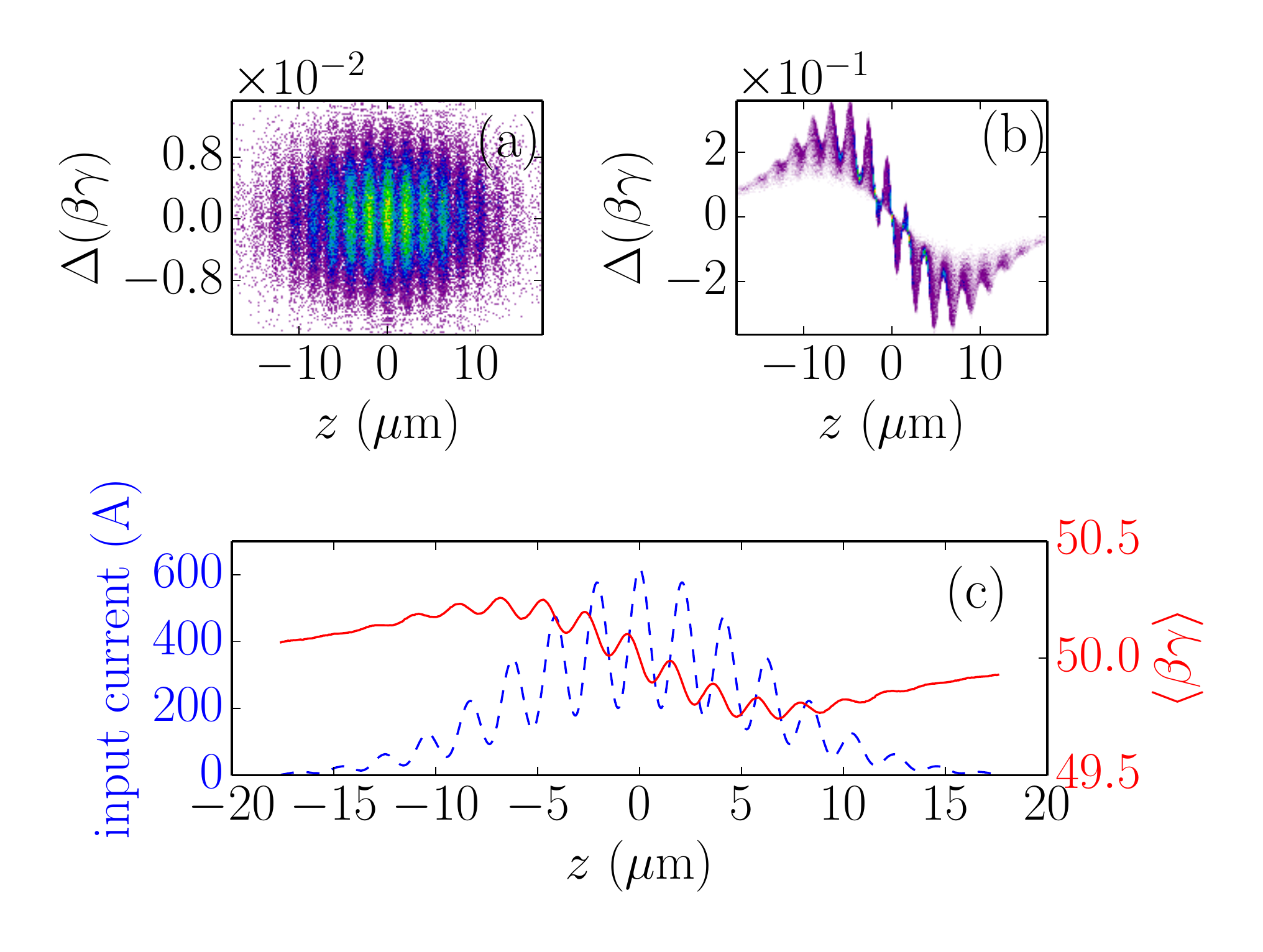} 
\end{center}
\caption{\label{phasespace} LPS distribution for a density-modulated Gaussian beam before (a) and after (b) the application of one space charge kick. Corresponding induced energy modulation (red solid trace) computed from image (b) and current distribution (blue dash trace) obtained from image (a); lower plot (c).In the upper images $\Delta(\beta\gamma)$ refers to the normalized momentum spread. }
\end{figure}

A comparison of the simulated LPS (after one space charge kick with {\sc elegant-bh}) with the initial one appears in Fig.~\ref{phasespace}(a,b) and demonstrates the salient features of the modulations and especially the $\pi/2$ shift between the final energy modulations and the initial density modulations; see Fig.~\ref{phasespace}(c).

\begin{figure}[hhhhh!!!!!]
\includegraphics[width=0.95\linewidth]{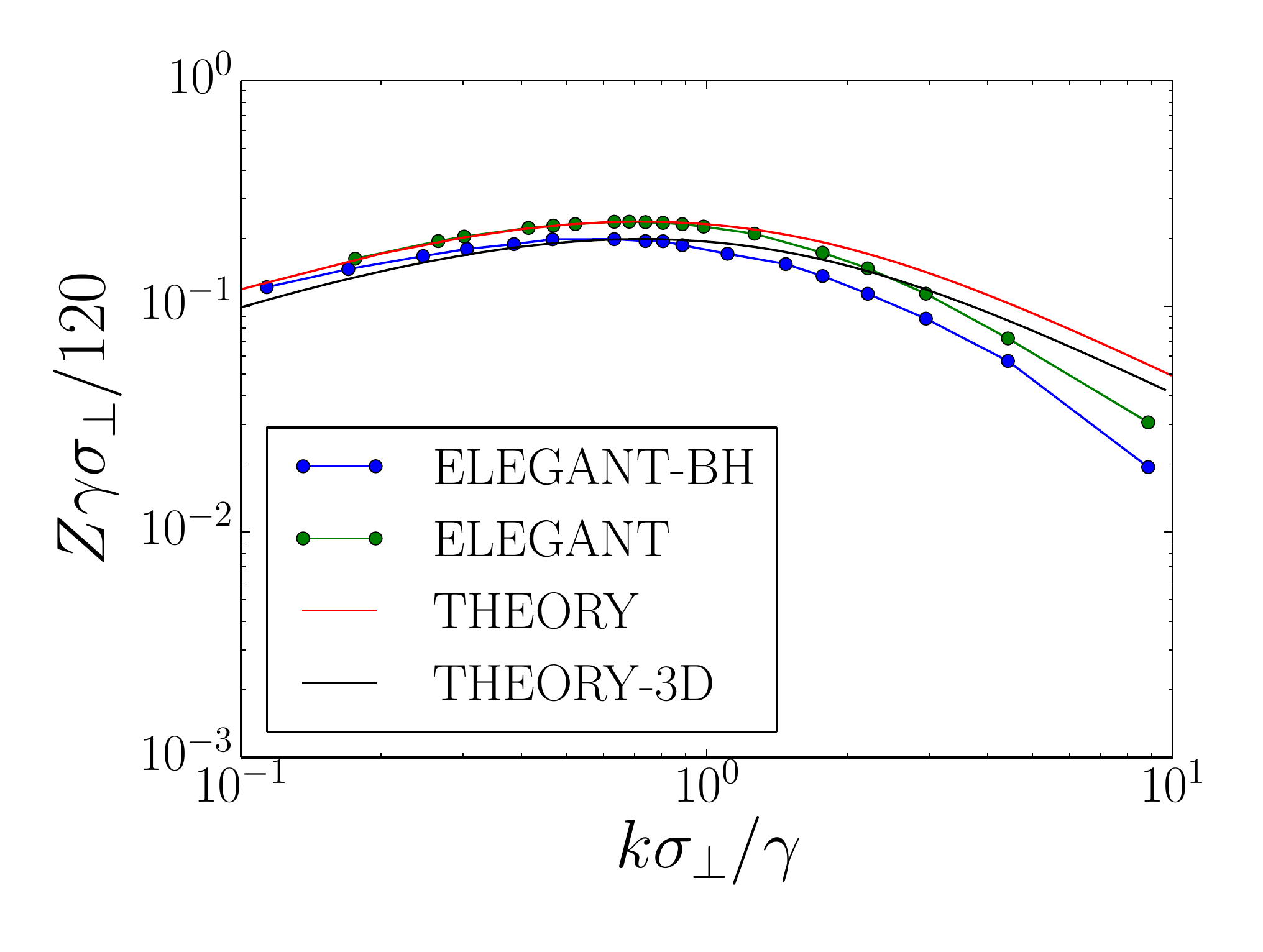} 
\caption{\label{impedance} Comparison of the longitudinal space-charge impedance computed from Eq.~\ref{imp_ven} (``THEORY"), from Eq.~\ref{imp3deq}) (``THEORY-3D"), with the ones retrieved from simulation with {\sc elegant} built-in {\tt LSCdrift} element (``ELEGANT"), and {\sc elegant-bh} (``ELEGANT-BH"). \label{fig:Z}}
\end{figure}

\begin{figure}[hhhhhhhh!!!!!!!]
\includegraphics[width=0.95\linewidth]{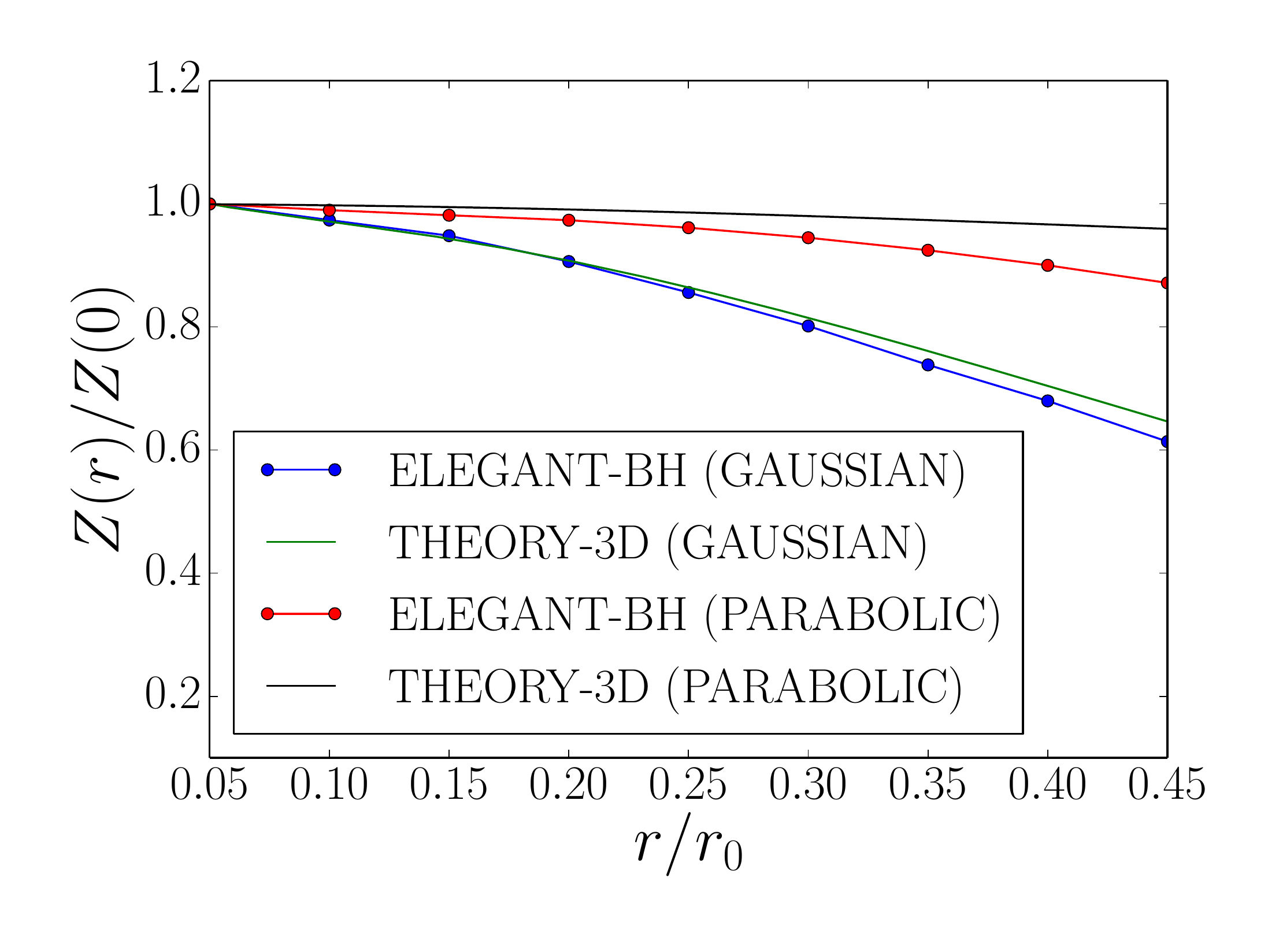} 
\caption{Radial dependence of  $Z(r)$ for a fixed value of $k$. The green and blue traces respectively correspond 
to the impedance with Gaussian transverse distribution from Eq.~\ref{imp3deq} and {\sc elegant-bh}. The parameter $r_0$ is the rms transverse size of the 
distribution.
\label{impedance_slice}}
\end{figure}

The Fourier transform of the initial current distribution was performed using a fast-Fourier transform (FFT) algorithm. Likewise, the mean energy of thin axial slices within the final LPS density distribution was computed thereby providing the energy modulation dependence on $z$, $\Delta {\cal E}_z(z)$. The extracted energy modulation $\Delta {\cal E}_z(z)$ is Fourier transformed to yield  $\Delta \widetilde{\cal E}_z(k)$ from which the axial electric field $E_z(k)$ was inferred.  The resulting impedance evolution as a function of $k$ obtained is compared against Eq.~\ref{eq:impedanceretr} in Fig.~\ref{fig:Z}.  The studies were carried out using different number of macroparticles ($N=[1,2,5,10]\times10^6$) 
to ensure the convergence and satisfy the statistical limit~\cite{Hirschmugl:1991, KJKim}. The number of FFT bins $n_b$ was also tuned to minimize discretization effects while varying  $k$. As the wavenumber $k$ value decreases, the bunch duration length was increased to ensure the number of  macroparticles per bin remains constant and guarantees a sufficient number of modulations occurs within the bunch. In our simulations we set this ratio to be typically $N/n_b\approx5000$. 
Figure~\ref{fig:Z}  points to small differences between the {\tt LSCdrift} element in {\sc elegant} that assumes transverse distribution to be uniform \cite{Borland:2000,Huang:2004} and Eq.~\ref{imp_ven}. The {\sc elegant-bh} performs full three-dimensional space charge force calculation and therefore inherits both the transverse and longitudinal effects in LSC impedance. 
Such effects were previously recognized~\cite{Juhao:2008} and are attributed to the radial dependence of the axial space charge field conferring a similar dependence on the impedance. To further explore longitudinal space charge impedance radial dependence we performed an analysis over thin radial slices $[r,r+\delta r]$ where $\delta r = 0.05 r_0$. 
The results of such an analysis are summarized in Fig.~\ref{fig:Z}, \ref{impedance_slice} for a beam following a transverse Gaussian distribution of the form $f(r) \propto \exp [-r^2/(2r_0^2)]$.
Figure \ref{impedance} compares the numerical results with the expected radial dependence analytically derived in Ref.~\cite{Juhao:2008}\footnote{We started with the Green's function for a $\delta$ ring derived as Eq. (35) of Ref.~\cite{Juhao:2008} and applied the Bessel recursive relation to further simplify Eq.(35) of the paper. 
In addition we explicitly wrote the $\mathbf {dr'}$ of Ref.~\cite{Juhao:2008}  as $2\pi r' dr' $ . }
\begin{eqnarray}\label{imp3deq}
Z(k,r) &=& \frac{4 \pi k}{\gamma^2} \int f(r') r 'dr' \left[\left(\theta(r-r') \right. \right.\nonumber\\
 & & \times K_0(\frac{k r}{\gamma})I_0(\frac{k r'}{\gamma})+  \nonumber\\ 
&&\left. \left. + \theta(r'-r)K_0(\frac{k r'}{\gamma})I_0(\frac{k r}{\gamma})\right) \right], 
\end{eqnarray}
where $\theta(r)$ is the Heaviside step function, $K_0$ and $I_0$ are modified Bessel functions, $f(r)$ is the distribution function. 
A noteworthy consequence of the observed strong radial 
dependence for a Gaussian transverse distribution is the effective smearing of the axial modulation which will effectively result
in weaker integrated energy modulations. 

As a final note, we point out that in the case of a parabolic $f(r)=f_0(a^2-r^2)\theta(r-a)$ and uniform $f(r)=f_0\theta(r-a)$ distribution 
an analytical form of the impedance can be retrieved [here $a$, $f_0$, and $\theta(r)$ are respectively the radius, normalization factor, and 
Heaviside function]. It is especially found that the parabolic transverse distribution yields an impedance with weaker dependence on the radius 
compared to a Gaussian transverse distribution; see Fig.~\ref{impedance_slice}. These observations suggest a possible use of transverse electron-beam shaping~\cite{maxonshaper,jiaoshaper} as ways of controlling the micro-bunching instability.

\section{LSCA simulations for FAST}

In this Section we explore the possible use of a staged LSCA beam line to produce micro-bunching structure with spatial scale corresponding to the ultraviolet regime, so $\lambda < 400$ nm. For our simulations we considered the configuration available at the Fermilab Accelerator Science \& Technology (FAST) facility (formerly known as ASTA)~\cite{asta} currently in its commissioning phase at Fermilab.  

\begin{figure}[hh!!!]
 \includegraphics[width=1\linewidth]{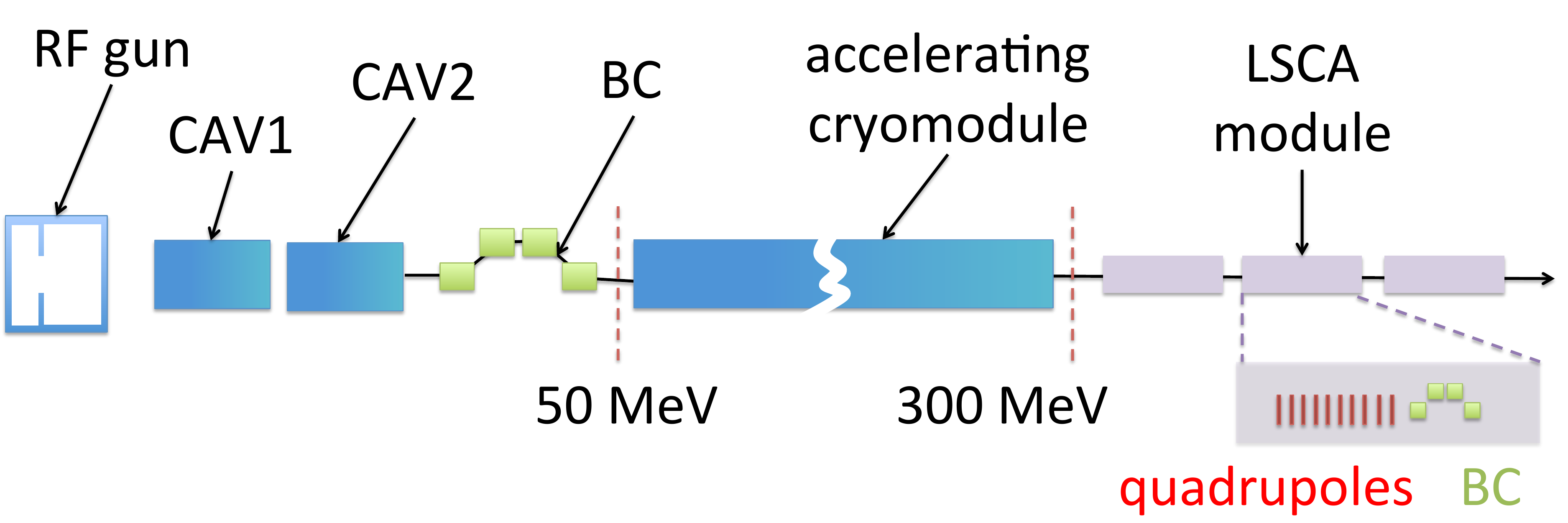}
\caption{\label{fig:fast} Overview of FAST facility and the proposed LSCA. The legend is as follows: "CAVx": accelerating cavities, "BC":
magnetic chicane bunch compressor, the thin (red) rectangles and (green) square symbols respectively represent the quadrupole and dipole magnets.  }
\end{figure}

The FAST facility is diagrammed in Fig.~\ref{fig:fast}. In short, the beam is produced from a  photocathode located in a 1+$\frac{1}{2}$  radiofrequency (RF) gun and accelerated to $\sim 50$~MeV by two superconducting TESLA accelerating cavities~\cite{Aune:2000gb}. Downstream of the accelerating cavities the beam can be manipulated (e.g. longitudinally compressed) and diagnosed in a 20-m transport line before being injected in an ILC type accelerating cryomodule composed of eight superconducting cavities. The beam, with final energies up to $\sim 300$~MeV, then can be directed into a small-footprint ring [the Integrable Optics Test Accelerator (IOTA)] or transported to 
experiments arranged along a $\sim 70$~m transport line. Conversely, the 70-m beamline, with proper optics, could accommodate the formation of broadband density modulation with UV spectral content; see Fig.~\ref{fig:fast}. \\
 
 \subsection{Initial beam parameters and LSCA beamline configuration} 

\begin{figure}[hh!!!]
 \includegraphics[width=1\linewidth]{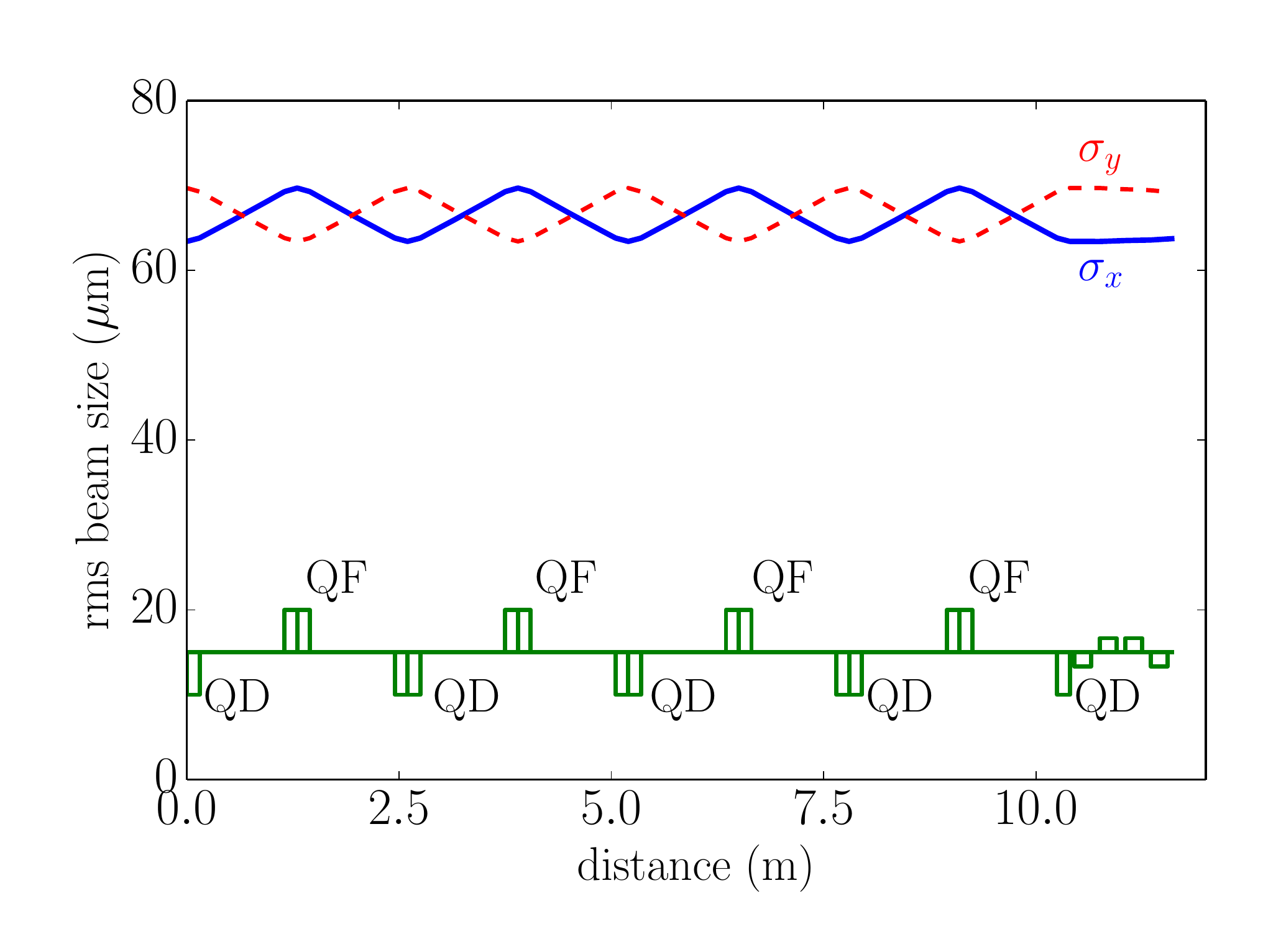}
\caption{\label{fig:lattice} Horizontal (blue) and vertical (red) rms beam size evolution in one LSCA module. The envelopes are obtained in the single-particle dynamics limit. 
The green diagram indicates the location of the focusing (QF) and defocusing (QD) quadrupole magnets while the four smaller rectangles (at distance $>12$~m) represent the chicane dipole magnets.}
\end{figure}

A numerical optimization of the electron-beam formation and acceleration to $\sim 50$~MeV was carried out with {\sc astra}~\cite{Piot:2010} for various charges. The results combined with
a mild bunch compression in the 50-MeV bunch compressor chicane, could produce bunches with 
peak current on the order of $\sim 500$~A and slice parameters gathered in Table~\ref{tab:Parameters}~\cite{Prokop:2013fz}. These parameters were used to generate initial distribution used in all the {\sc elegant-bh} simulations presented below. For simplicity we consider all the LSCA modules to be similar: they consist of 4 FODO-cell sections each followed by small bending angle chicanes. The only difference between the modules is the $R_{56}$ parameters associated to the chicanes as explained below. The horizontal dispersion introduced by the chicanes is minimal and does not break the periodicity of the FODO cells. The settings of quadrupole magnets arranged as a FODO cell were optimized using single-particle dynamics simulations. The corresponding beam size evolution along one LSCA module appears in Fig.~\ref{fig:lattice}.
%
%%%%%%%%%%%%%%%%%%%%%%%%%%%%%%%%
%
\begin{table}[ht!]
\caption{\label{tab:Parameters} Beam parameters considered for the LSCA simulations using the 
setup of Fig.~\ref{lsca-layout}. }
\begin{center}
\begin{tabular}{l c c}
\hline
{\bf Parameter, Symbol} & {\bf Value} & {\bf Units}\\
\hline  
   Transv. spot size, $\sigma_{x,y}$ & 68.0 & $\mu$m  \\
   Charge, $Q$ & 20.0 &  pC   \\
   Lorentz factor, $\gamma$ & 600 & --   \\
   RMS bunch duration, $\tau$ & 19 & fs \\
   Peak current, $I$ & 415 & A\\
   Transv. emittance, $\varepsilon_{x,y}$ & $5\times 10^{-8}$ & m \\
   Frac. momentum spread, $\sigma_{\delta}$ & $10^{-4}$ & -- \\
   Number of macroparticles, $N$ & $[10^6,10^7]$ & -- \\
   %Total LSCA length, $D$ & 28.0 & m \\
\hline 
\end{tabular}
\end{center}
\end{table}

In our simulation we model the evolution of a thin longitudinal slice of the bunch with an equivalent peak current and slice parameters listed in Table~\ref{tab:Parameters}. The slice is assumed to have a longitudinal Gaussian distribution (to avoid complication arising from sharp transition of a uniformly-distributed slice). The transverse distribution is also chosen to be Gaussian along the horizontal ($x$) and vertical ($y$) axis. 
Likewise, the divergence and energy distribution are all taken to be Gaussian.

\subsection{Optimization of one LSCA module} 

\begin{figure}[hhhhh!!!!!]
\centering
\includegraphics[width=0.98\linewidth]{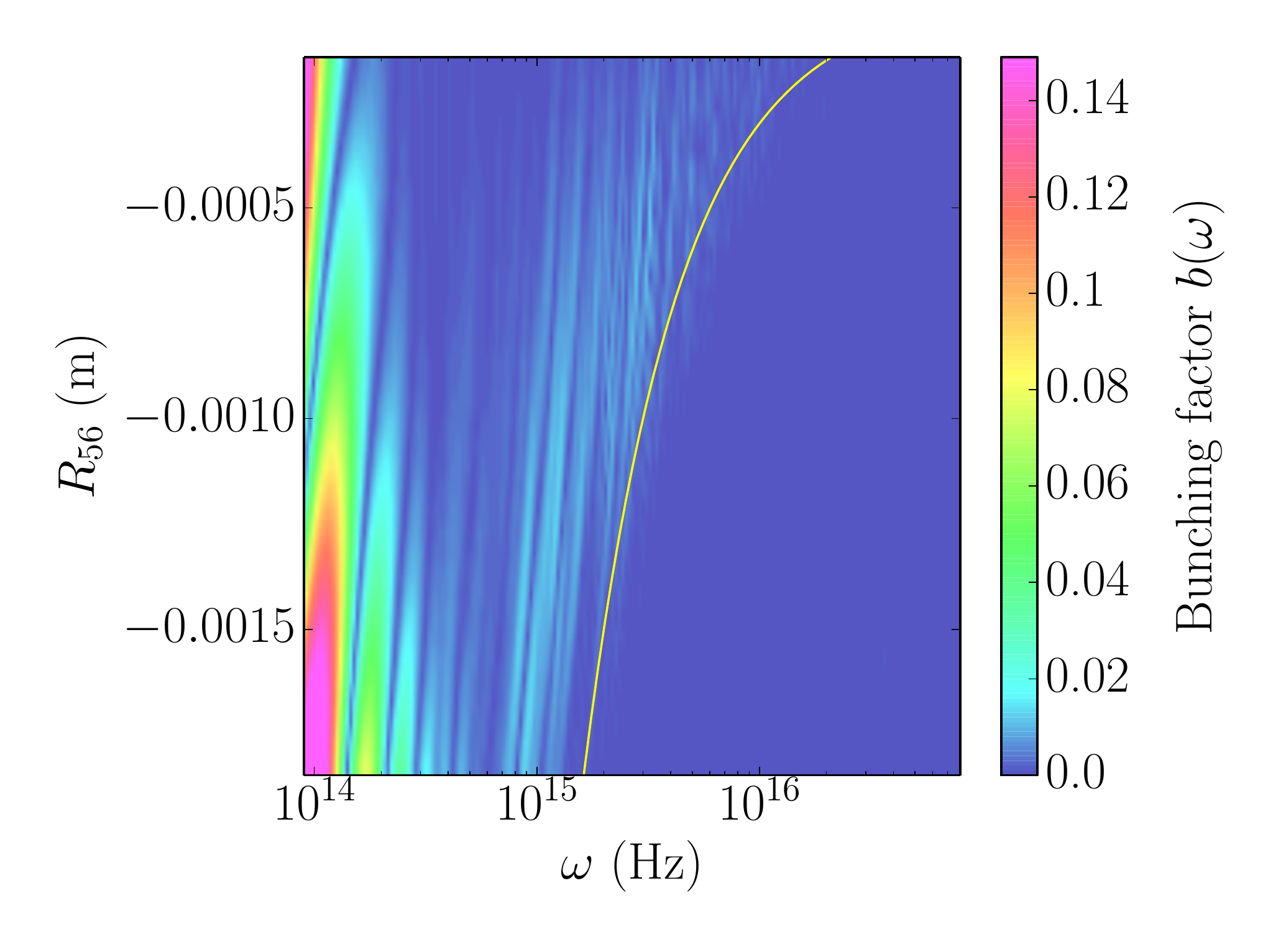}
\caption{\label{R56scan}The evolution of the bunching factor $b(\omega)$ as a function of longitudinal dispersion $R_{56}$ and the modulation 
wavelength $\omega$ after
passing through FODO+BC. The superimposed yellow trace represents the exponential cut-off described by Eq.~\ref{eqn:cutoff}.}
\end{figure}

We start with the optimization of one LSCA module consisting of several FODO sections and one BC. We varied 
two parameters at this point: the quadrupole magnet distance in the FODO sections and the bending angle in the chicane which 
affects its longitudinal dispersion $R_{56}$. As the goal of this study is to reach the shortest wavelength possible at FAST, 
we focus on small $R_{56}$ values. Figure~\ref{R56scan} provides the bunching factor in the frequency 
range $\omega \in [1\times 10^{14}, 5\times 10^{16}]$~Hz for varying value of $R_{56} \in -[2, 0.1]$~mm. Figure.~\ref{R56scan} also indicates
that the high-frequency content suppression is following the expected scaling of Eq.~\ref{eqn:cutoff}. For the considered case of
betatron function and reachable energy the optimum wavelength is $\lambda_{opt}\approx 750$~nm 
(corresponding to $\omega_{opt} \approx 2.5\times 10^{15}$~Hz), the broadband feature of the amplification process 
has spectral content up to $\lambda \sim 190$~nm (corresponding to $\omega\sim 1\times 10^{16}$~Hz). 

\begin{figure}
\centering
\includegraphics[width=0.95\linewidth]{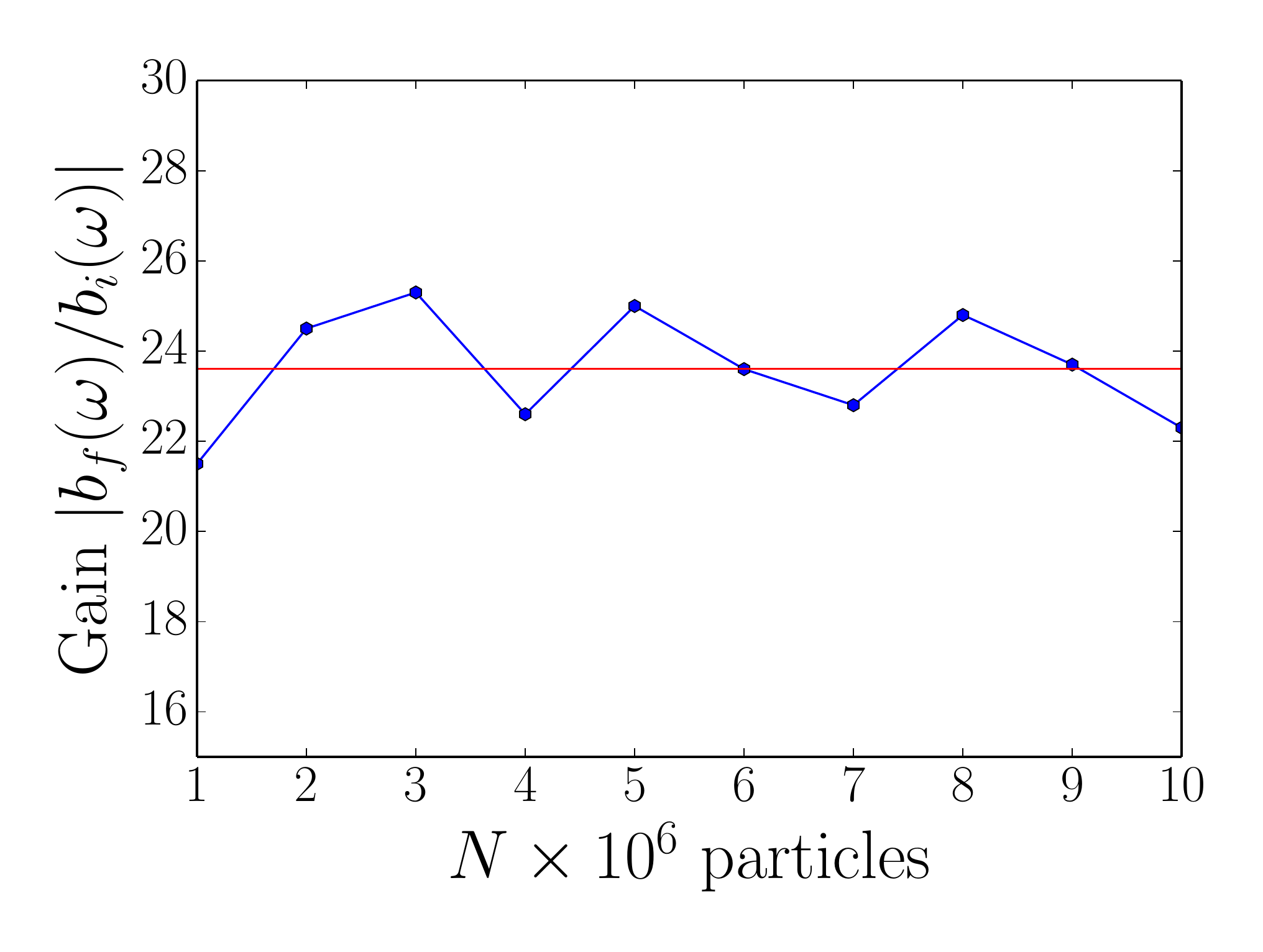}
\caption{\label{gainN} Gain for the first LSCA module as function of number of macro-particles $N$ representing the beam (filled circles). The selected longitudinal dispersion and associated optimum wavelength are respectively $R_{56}=364$~$\mu$m, and $\lambda_{opt} \simeq 750$~nm. The red trace represents the mean gain value obtained by averaging over the 10 simulation sets. }
\end{figure}

The {\sc elegant-bh} simulations presented in Fig.~\ref{R56scan} were performed with $N=10^7$ macro-particles and $\Delta R_{56}=5\mu$m (while 
the slice actually contains $N_e=125\times 10^6$ electrons). Therefore
the noise floor \cite{Hirschmugl:1991,KJKim} of the bunching factor is  $\simeq 1/\sqrt{N_e}\simeq 9 \times 10^{-5}$ while our 
simulations are limited to noise floor of  $\simeq 1/\sqrt{N}\simeq 3 \times 10^{-4}$. To verify the limited number 
of macro-particles does not significantly affect the retrieved gain we carried out numerical simulation for different 
values of $N$ and found no dependence as seen in Fig.~\ref{gainN}. In the latter case the longitudinal dispersion was 
set to $R_{56}=364$~$\mu$m corresponding to an optimum wavelength of $\lambda_{opt} \simeq 750$~nm. The gain averaged 
over the 10 simulation is $\mean{G}=23.6$ and its standard deviation $\mean{G^2}^{1/2} = 1.28$ corresponding to a fractional spread 
of  $\sim 6$~\%. These results essentially demonstrate that our gain-calculation technique is independent of the number of macroparticles used in the simulation. 

\begin{figure}
\includegraphics[width=0.98\linewidth]{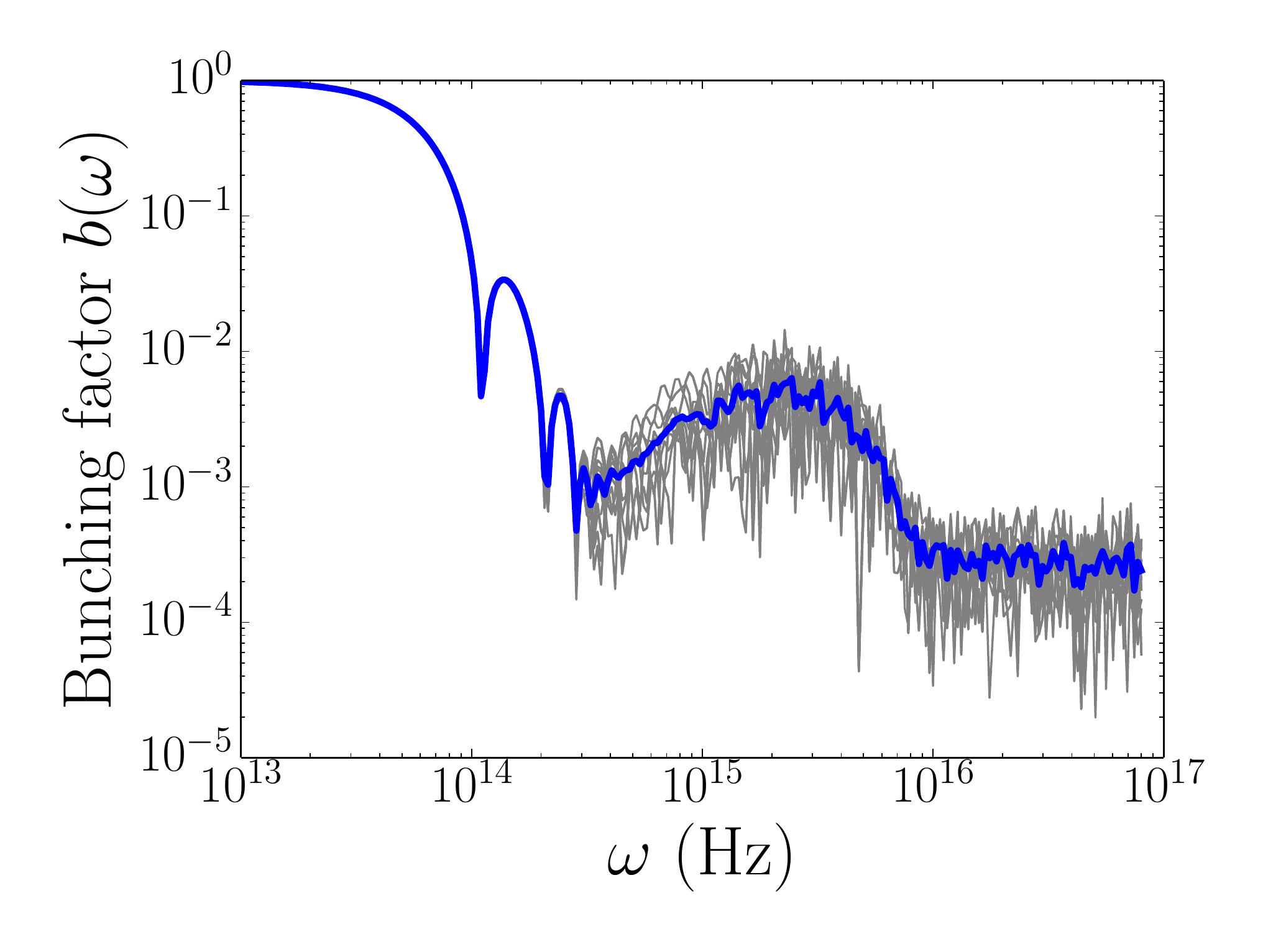}
\caption{\label{averaging} Bunch factor evaluated downstream of one LCSA module with $R_{56}= 364$~$\mu$m. The gray traces represent the bunching factors computed for 20 independent realizations of the initial bunch distributions while the solid blue trace corresponds to the averaged value. }
\end{figure}

\begin{figure}[hh!!!]
 \includegraphics[width=1.0\linewidth]{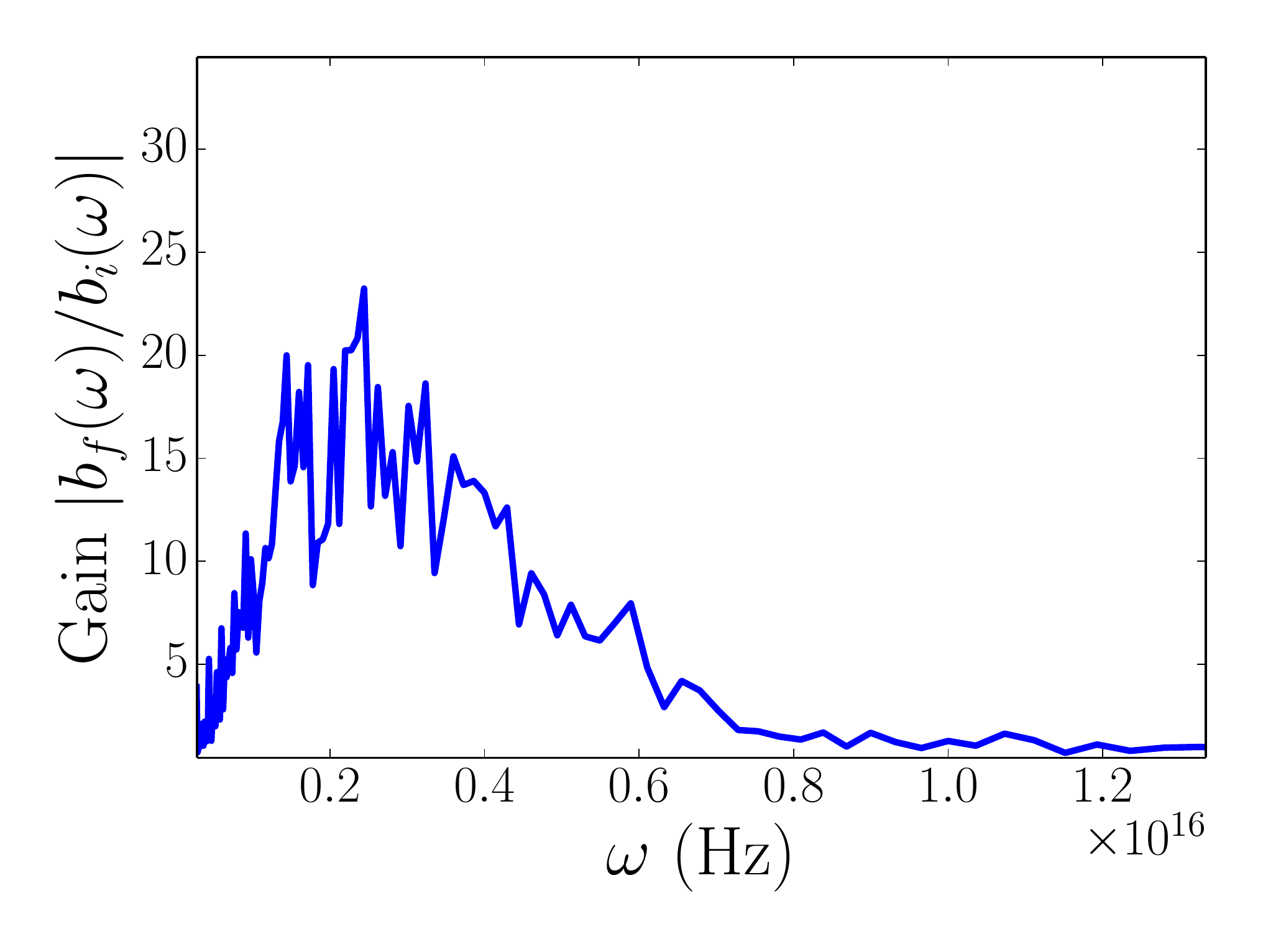}
\caption{\label{Gain} Gain curve as a function of frequency in the interval where significant gain is obtained. The curve
is computed for a single (the first) LSCA module.}
\end{figure}

The striations observed along the frequency axis in Fig.~\ref{R56scan} arises
from the initial shot-noise~\footnote{The same random realization of the initial distribution was used while scanning the value of $R_{56}$.}. The LSCA process, being seeded by the initial shot-noise in the beam, 
fluctuates shot-to-shot as different beam distribution is realized. Consequently, to perform noise-insensitive bunching factor analysis, we carried out
20 {\sc elegant-bh} runs for given lattice settings with different initial random seeds in order to generate independent realizations of initial
bunch distribution. As an example we evaluated the bunching factor computed over the spectral region of interest $\omega \in [1 \times10^{13}, 9\times 10^{16}]$~Hz statistically averaged 
over the 20 independent runs appears in Fig.~\ref{averaging} (blue trace) and demonstrates that, in average, the frequency 
in the range $\omega \in [3 \times10^{14}, 1\times 10^{16}]$~Hz is associated with enhanced value of the bunching factor. 
Likewise, an average gain curve over the region of interest can be computed; see Fig.~\ref{Gain}. In the case of a superconducting linac operating in a burst mode this type of average gain curve will practically be generated over a single burst (corresponding to a 1-ms RF macropulse accelerating 3000 statistically-independent bunches in the case of the FAST facility). 

%

%The total length of the LSCA was chosen to be 28 m.
%{\color{red} what happened to the length scan?}
%

The latter figure reports the gain computed as the ratio between the final and initial bunching factors $|b_f(\omega)/b_i(\omega)|$.  
To smooth out the shot-to-shot nature of the gain, the presented gain is averaged over 20 random realizations of the initial macro-particle 
distribution. 

As a concluding remark, we note that by introducing a chirp $C$ it is possible to compress the optimum modulation wavelength to shorter value given by $\lambda_{comp} \simeq \lambda_{opt}/\kappa$ where the compression factor $\kappa$ was introduced above.

\subsection{Cascaded LSCA} 

To simulate a 3-stage LCSA module, we iterated the process described in the previous section for each stage 
so to ensure the $R_{56}$ is properly optimized. The simulations were carried out in a piecewise fashion. First, the FODO channel of stage $n$ was simulated with space charge, the output was passed to the subsequent BC. The $R_{56}$ was optimized to provide the largest bunching factor at the selected wavelength.  
The resulting distribution was rematched and then passed to the $n+1$ FODO channel where the process was repeated. 
As mentioned earlier the chicane have small $R_{56}$ and single-particle dynamics does not affect the matching. 
However, in the presence of space charge for a 300-MeV beam, we find that the matching is  
deteriorated therefore requiring rematching of the beam parameters after each module. The final optimized 
values for the $R_{56}$ for first, second and third stage are respectively $-364$, $-279$, and $-142$~$\mu$m. 

The total length of cascaded LSCA configuration is $28$~m. This distance was selected based on the FAST lattice 
parameters and the amount of energy modulation acquired for one FODO cell. We note that other configurations are also possible (e.g. with longer drifts). 

\begin{figure} 
\centering
\includegraphics[width=1.\linewidth]{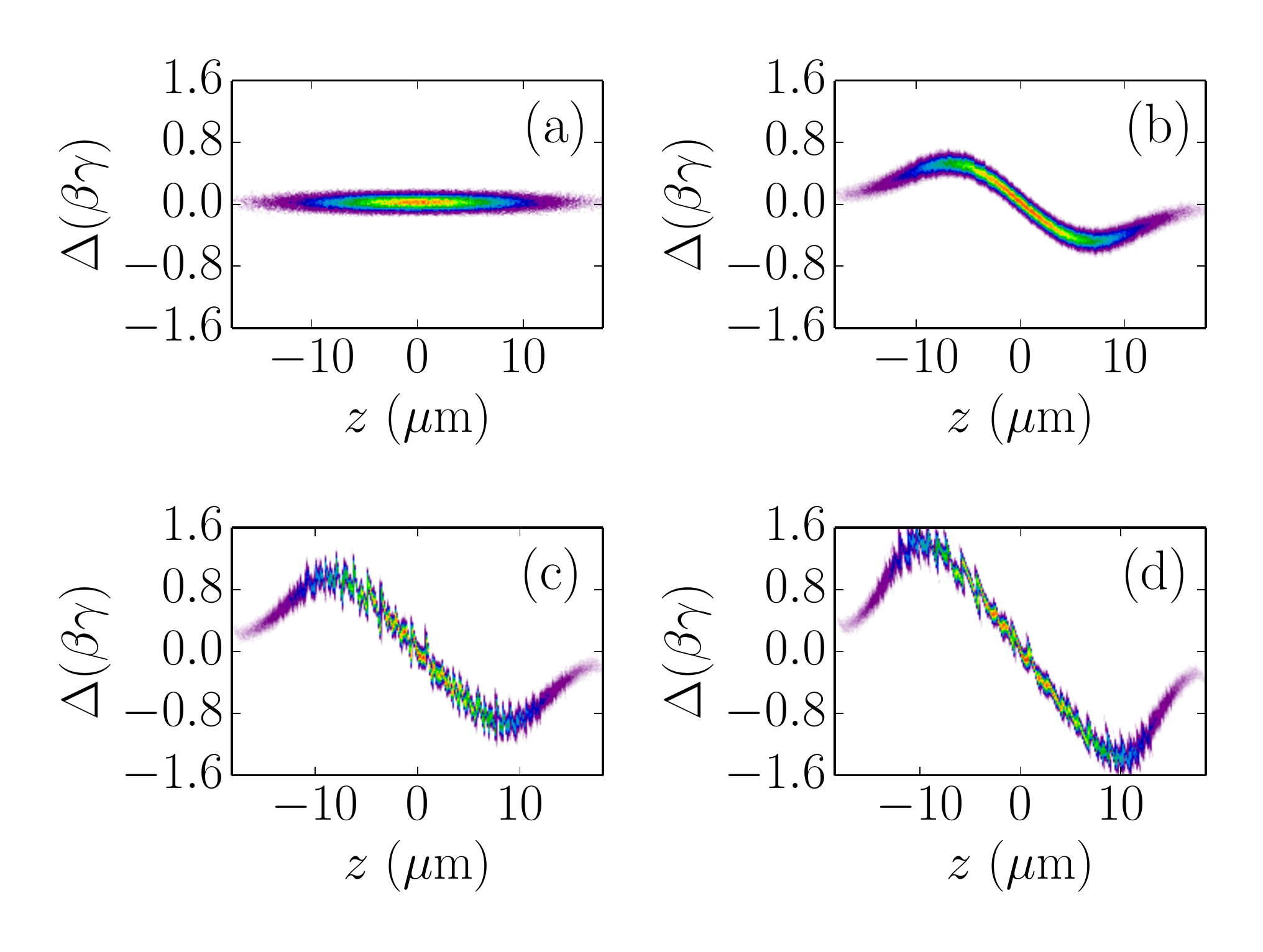} 
\caption{\label{phasespaceev} Snapshots of  LPS evolution along the cascaded LSCA: initial (Gaussian) bunch before (a) and after passing through
one (b), two (c) and three (d) LSCA modules. The Lorentz factor is $\gamma=600$ and $10^7$ macroparticles were used in these simulations. }
\end{figure}
\begin{figure}[hhhhhh!!!!]
\includegraphics[width=0.99\linewidth]{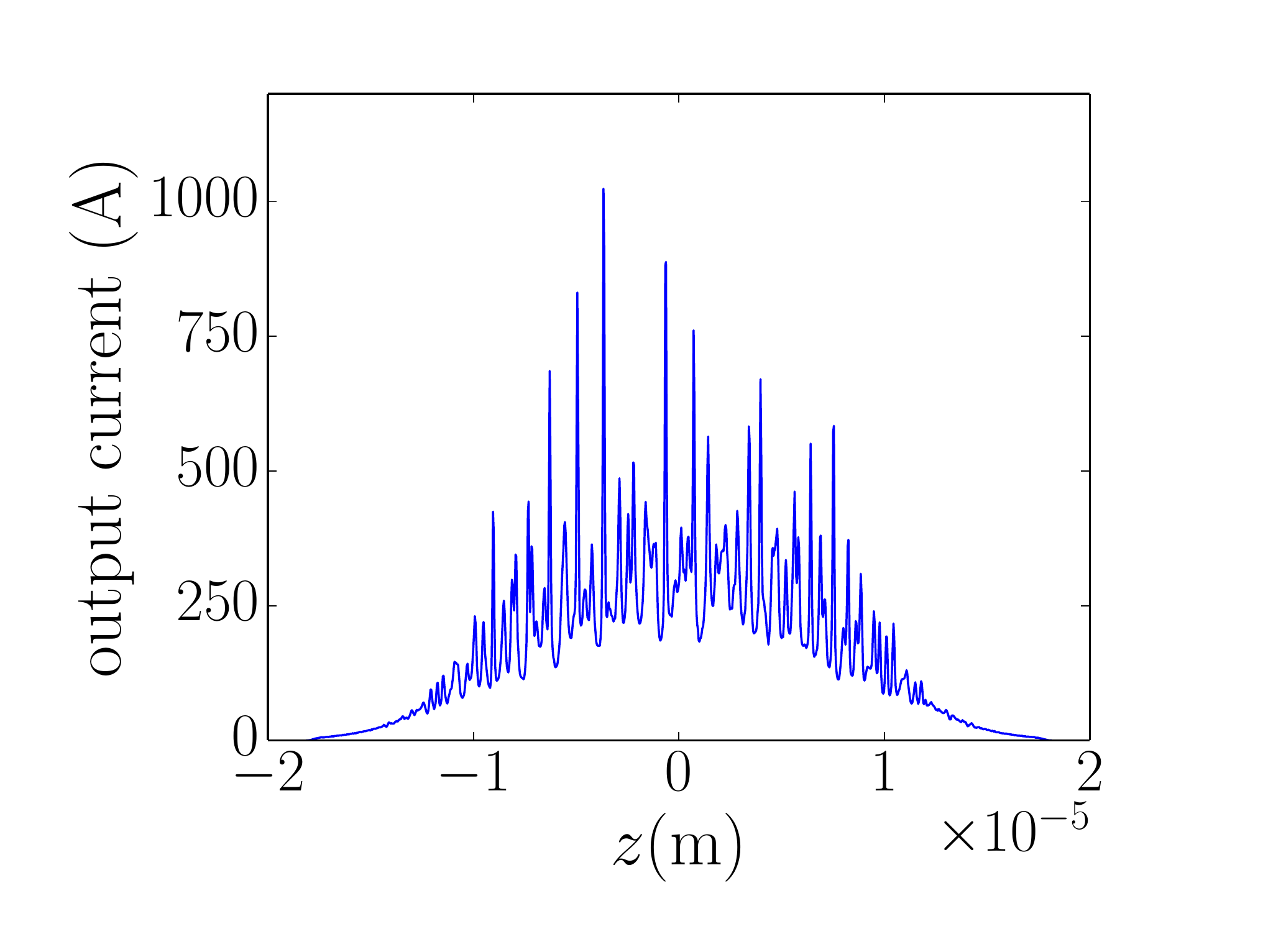}
\caption{\label{outcurrent} Current profile upstream (red dashed trace) and downstream (blue solid trace) of the LSCA. 
For these simulations the number of macro-particles was taken to be $N=10^7$. }
\end{figure}

Evolution of the LPS associated to the 500-A slice being tracked throughout the LSCA modules appears in Fig.~\ref{phasespaceev}. At $\gamma = 600$ strong density modulations start to form downstream of the second LSCA module. The shortest microstructures in the LPS are achieved as the LSC-induced modulation has its local correlation $C_{loc}$ satisfying $R_{56}=-1/C_{loc}$ giving rise to structured density profile with shortest temporal scale on the order of 
$\tau \simeq  R_{56} \sigma_{\delta, u} /c$. Eventually the LPS becomes strongly disrupted as seen in Fig.~\ref{phasespaceev}(d).

The microbunches have durations of the order of hundred femtoseconds and could in principle lead to the generation of attosecond radition pulse; e.g. when co-propagated with ultrashort laser pulse in an undulator; see Ref.~\cite{Dohlus:2011}. Because of the local energy chirp, additional effects related to wakefields and other interactions with radiation [e.g. coherent synchrotron radiation (CSR)] should also be  taken into account. 

\begin{figure}[ht]
\includegraphics[width=0.94\linewidth]{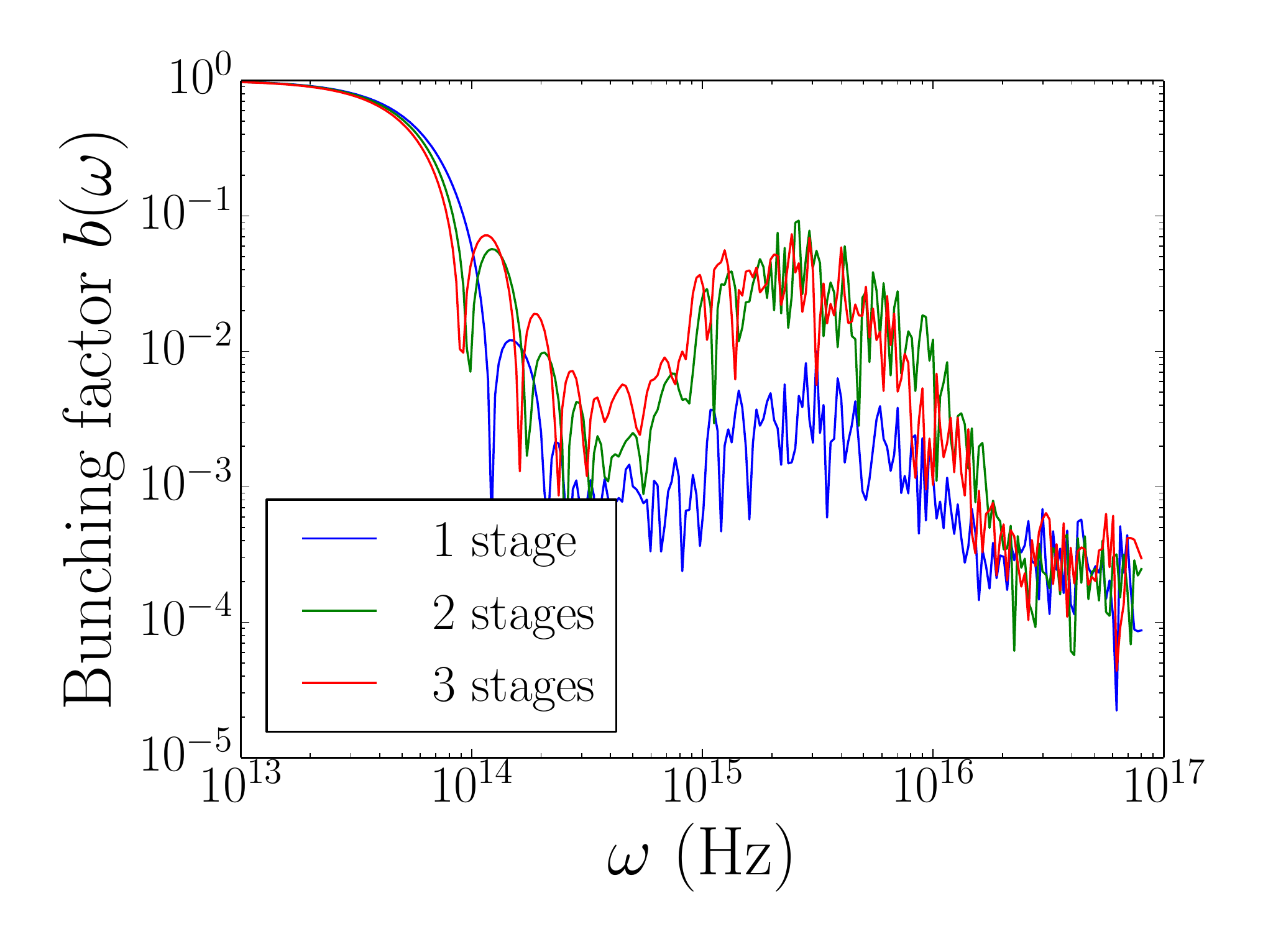}
\caption{\label{bfactor} Bunching factor extracted from the LPS's shown in Fig.~\ref{phasespaceev}. The blue, red, and green traces respectively correspond the the bunching factor downstream of the first, second and final LSCA stages.  }
\end{figure}

Figure~\ref{bfactor} represents the evolution of the bunching factor after each LSCA stage averaged over 20 random realizations of the initial bunch distribution. The broad spectral features of the bunching factor observed in Fig.~\ref{R56scan} are preserved until the end of the final stage. From the evolution of the bunching factor we inferred a total gain of $G\approx 500$ (evaluated at the optimum wavelength) for the considered three-stage LSCA . 

One limitation found in the present study is the cumulated energy spread which leads to transverse emittance growth via chromatic aberrations. This emittance dilution eventually leads to the suppression of the modulation (via an angular smearing  effect). 
Overall, this effect results in saturation of the gain in the final stages as seen in Fig.~\ref{bfactor}.  We actually find that gain for the first and second stage is $G\approx 20$ while it is only $G\approx  1.3$ for the last stage. \\

\begin{figure}[ht]
\centering
\includegraphics[width=0.98\linewidth]{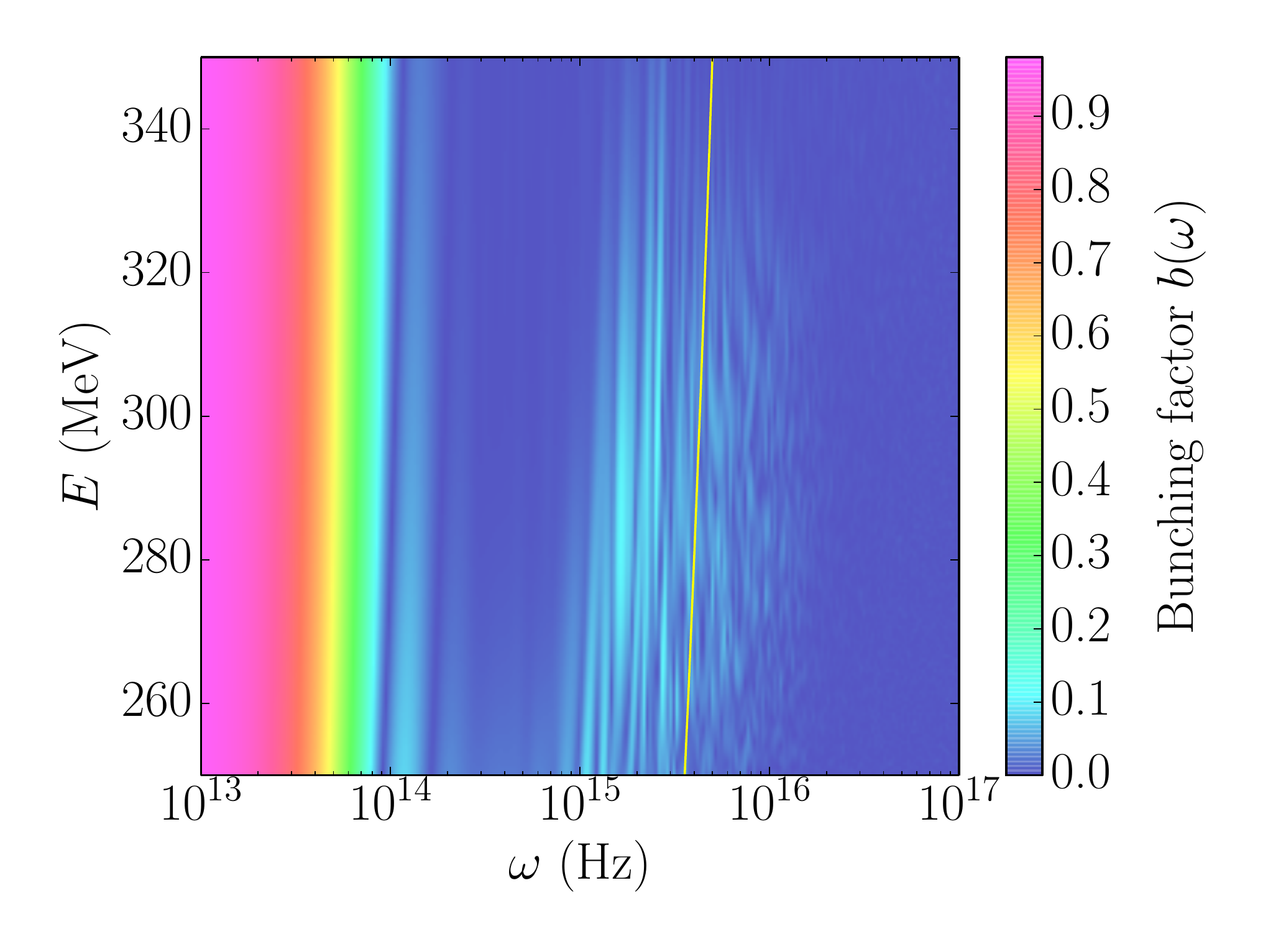}
\caption{\label{fig:energyscan}Evolution of the bunching factor $b(\omega)$ as a function of bunch energy. The superimposed yellow line represents
the optimal wavelength $\lambda_{opt}$.
}
\end{figure}

We finally point out that increasing the energy would provide a path to shorter wavelength as expected from the scaling described by Eq.~\ref{eqn:lambdaopt}. Such an opportunity is depicted in Fig.~\ref{fig:energyscan}. Again the simulation qualitatively captures the expected scaling for the optimum frequency $\omega_{opt}\equiv 2\pi c/\lambda_{opt}$ (though the striations prevent from a quantitative comparison).  

\subsection{Compressed case}
\begin{figure}[hhhh!!!!!!!]
 \includegraphics[width=1.\linewidth]{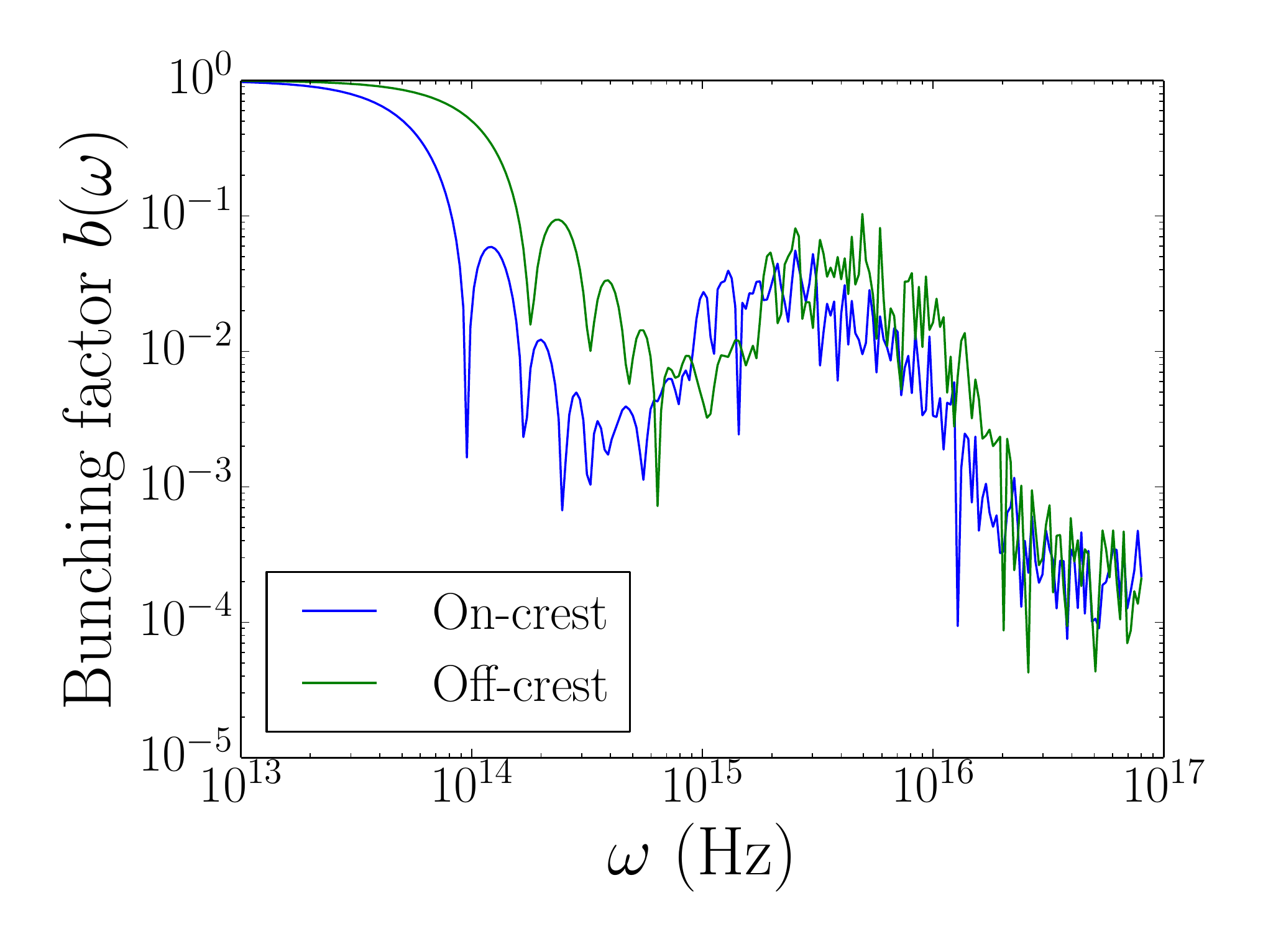}
\caption{\label{3stage} Simulated bunching factor downstream of the third LSCA stage for the case of an uncompressed (blue) and compressed (green) incoming beam. The simulations are performed with $10^7$ macroparticles.}
\end{figure}

In the previous sections we specialized to the case where the incoming LPS is uncorrelated. Introducing a LPS chirp can significantly decrease the wavelength to lower values well in the ultraviolet (UV) range. 
As an example, we computed the final bunching factor obtained from the same setup as in the previous section
with an initial LPS with a chirp ${\cal C} \equiv \frac{d\delta}{dz}\big|_0=1667$~m$^{-1}$. This chirp is numerically 
applied and corresponds to the value that would result in a maximum compression downstream of the three BCs used in the LSCA module. 
The resulting bunching factor has significant content ($b_f \simeq 1$\%) at $\lambda\approx140$~nm; see Fig.~\ref{3stage} (green trace). 

Here we stress that, for simplicity, the chirp was ``numerically applied'' just before the last 
bunch compressor (thus its large value). In practice such a large chirp might be challenging to achieve using conventional 
off-crest operation of the linac (especially given the low operating frequency of the considered superconducting linac 1.3 GHz). 
However, 11.4 GHz X-band linac with an energy gradient of 100 MeV/m operated at zero crossing can provide a required chirp within 4 m. 
We also note that advanced techniques such as the application of a nonlinear chirp 
with a dielectric-lined waveguide might provide the required chirp within shorter length~\cite{francois}. Additionally, milder compression might be implemented (corresponding to much lower chirps). 
 
 \subsection{Radiation mechanism} 
The electromagnetic radiation  emitted by a bunch of electrons has its spectral-angular fluence given by
\begin{eqnarray}
\frac{dW}{d\omega d\Omega}=[N + N(N-1) b(\omega)^2] \frac{dW}{d\omega d\Omega}\big|_1,
\end{eqnarray}
 where $\frac{dW}{d\omega d\Omega}\big|_1$ represents the single-electron radiation spectral fluence associated to the considered electromagnetic process.  The latter equation assumes the beam follows line-charge distribution. In practice when the beam has a transverse extent the radiation is suppressed $\frac{dW}{d\omega d\Omega} \rightarrow \frac{dW}{d\omega d\Omega} \times {\cal F} $ where a multiplicative suppression factor (${\cal F} \le 1$) has to be included~\cite{saldinsuppression}. In principle any radiation mechanisms can be considered. Here we consider the case when the beam simulated in the previous section is passed through an undulator magnet thereby generating undulator radiation (UR). The fundamental wavelength of UR is related to the undulator period $\lambda_u$ via  
 \begin{eqnarray}
 \lambda = \frac{\lambda_u}{2\gamma^2} \left( 1+ \frac{K^2}{2} + \gamma^2\theta^2 \right), 
 \end{eqnarray}
 where $\theta$ is the direction of observation with respect to the electron-beam direction 
 and the undulator parameter is $K\equiv \frac{eB}{k_u mc}$, where $B$ is the undulator peak field, $k_u \equiv \frac{2\pi}{\lambda_u}$, and $e$, and $m$ are respectively the electronic charge and mass. In order to reach ultraviolet wavelength on axis ($\theta=0$) for $\gamma = 600$ we select an undulator period of $\lambda_u=5$~cm and a tunable $K \in [0.49, 3.9]$. The latter parameters correspond to the {\tt U5.0} ALS undulator~\cite{ALSund50}. The range of attainable undulator parameters would allow for radiation to be generated within the spectral range $\lambda \in [78, 498]$~nm which covers the range where LSCA-induced micro-bunching is sustained.

\begin{figure}[hhhh!!!!!!!]
 \includegraphics[width=1.\linewidth]{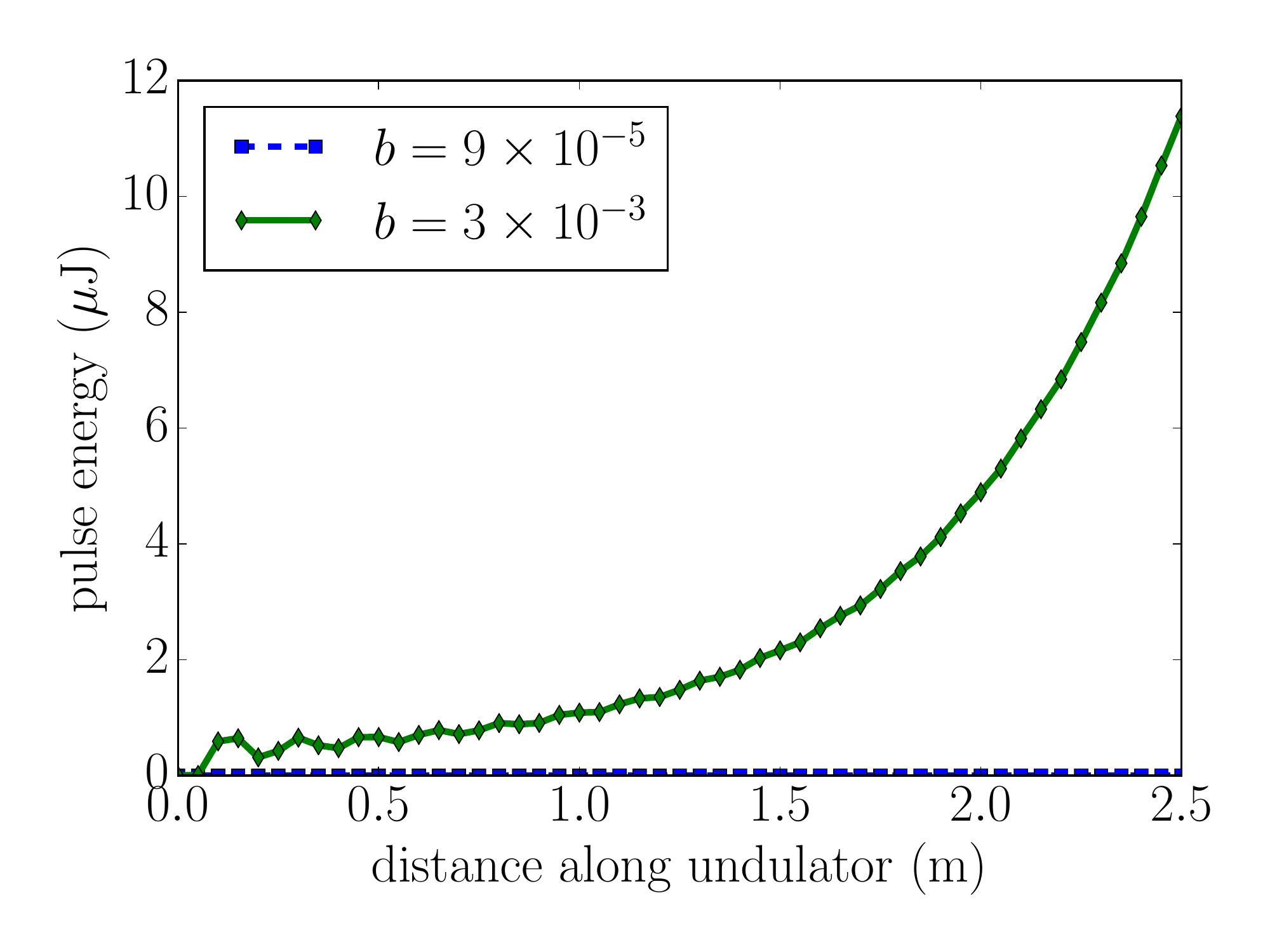}
\caption{\label{fig:Radiation} Evolution of the undulator radiation for $\lambda= 235.5$~nm for a scaled bunching factor of $3\times 10^{-3}$ (blue trace, diamond symbols). The radiation produced by a beam without micro-bunching modulations, e.g. with bunching factor $b=1/\sqrt{N_e} \simeq 9\times 10^{-5}$, is also displayed (green trace, square symbols)  for comparison.}
\end{figure}

\begin{table}[hhhh!]
\caption{Electron-beam and undulator parameters used for the {\sc genesis} simulations. \label{tab:FELinput} }
\begin{center}
\begin{tabular}{l c c}
\hline
{\bf Parameter, Symbol} & {\bf Value} & {\bf Units}\\
\hline  
   Transv. spot size, $\sigma_{x,y}$ & 50.0 & $\mu$m  \\
   Lorentz factor, $\gamma$ & 600 & --   \\
   Peak current, $I$ & 500 & A\\
   Transv. emittance, $\varepsilon_{x,y}$ & $5\times 10^{-8}$ & m \\
   Frac. momentum spread, $\sigma_{\delta}$ & $5\times 10^{-4}$ & -- \\
\hline 
   Undulator period, $\lambda_u$ & 5.0 & cm  \\
   Undulator parameter, $K$ & 2.18 & --  \\
   Radiation wavelength, $\lambda$ & 235.5 & nm  \\
\hline 
\end{tabular}
\end{center}
\end{table}

To quantitatively assess the properties of undulator radiation produced seeded by a LSCA-microbunched beam, we employ the program {\sc genesis}~\cite{genesis}. The beam parameters are consistent with the beam parameters generated downstream of the LCSA; see Tab.~\ref{tab:FELinput}. For our simulation we rescale the bunching factor obtained with $N=10^7$ macroparticles, see Fig.~\ref{3stage}, to the real number of electron in the considered axial slice $N_e=1.2\times 10^8$ following $b(\lambda) \rightarrow \sqrt{\frac{N}{N_e}} b(\lambda)$ where $\sqrt{\frac{N}{N_e}} \simeq 0.29$. This scaled bunching factor obtained for a specific wavelength is then used as an input in the steady state {\sc genesis} simulation. The simulated evolution  of the radiation pulse energy along the undulator length appears in Fig.~\ref{fig:Radiation}. In the latter figure we considered a wavelength of  $\lambda= 235.5$~nm with a scaled bunching factor of $\simeq 3\times 10^{-3}$ (i.e. a nominal bunching factor of $1\times 10^{-2}$ in Fig.~\ref{3stage}). Our simulations  demonstrate that UV pulses with energies on the order of $\sim 10$~$\mu$J could be reached downstream of a meter-scale section of the undulator three orders of magnitudes above the ``shot-noise" radiation of 9 nJ (achieved with an initial shot-noise bunching factor of $b=1/\sqrt{N_e}$). 

\section{Summary}
Using a grid-less code adapted from Astrophysics we have investigated three-dimensional effects in the LSC (longitudinal space-charge)
impedance and confirmed that the one-dimensional often used LSC impedance model is a good approximation. 
Additionally we benchmarked the radial dependence of the LSC impedance with the analytical results developed in Ref.~\cite{Juhao:2008}. 

Finally, we confirmed the possible use of a cascaded LSCA (longitudinal space-charge amplifier) scheme to produce femtosecond 
microstructures in the LPS (longitudinal phase space) with spectral range attaining the ultraviolet 
domain using electron-beam parameters  achievable at FAST facility.  

\section{Acknowledgments}
We are grateful to Dr. J. Barnes (U. Hawaii) for granting us the use of his open-source version of the BH algorithm and to Dr. M. Borland (ANL) for his help with {\sc elegant}. We thank Dr. J. Wu (SLAC) for discussing some of the results presented in his paper~\cite{Juhao:2008}, and Dr. R. Li (JLab) and Dr. V. Litvinenko (BNL) for their interests and comments on our work. This work was supported by the US Department of Energy (DOE) under contract DE-SC0011831 with Northern Illinois University. Fermilab is operated by the Fermi Research Alliance LLC under US DOE contract DE-AC02-07CH11359.

\appendix
\section{Space charge algorithm validation \label{app:benchmarks}}
To gain confidence in the algorithm developed we benchmarked our simulations in the macroscopic regime and rely on both analytical results and simulations carried out with the {\sc astra} program \cite{ASTRAmanual}. 

\begin{figure}[hhhh!!!!]
\includegraphics[width=0.96\linewidth]{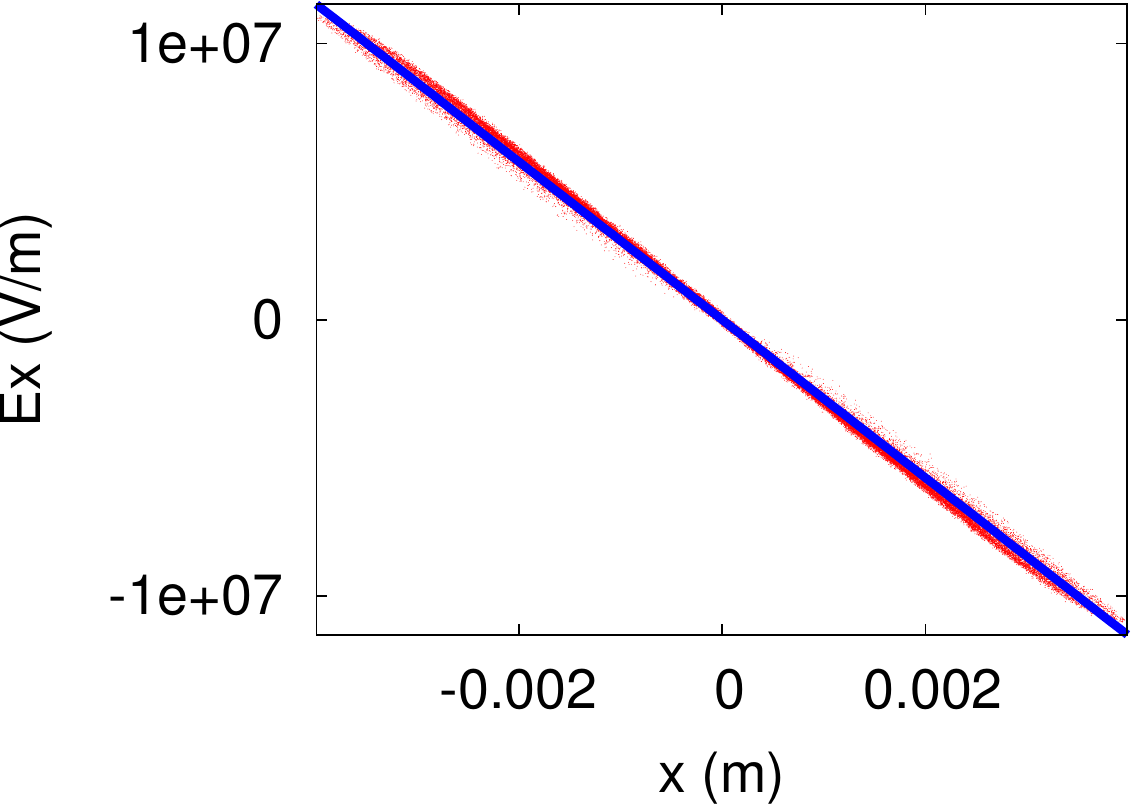}\\
\includegraphics[width=0.96\linewidth]{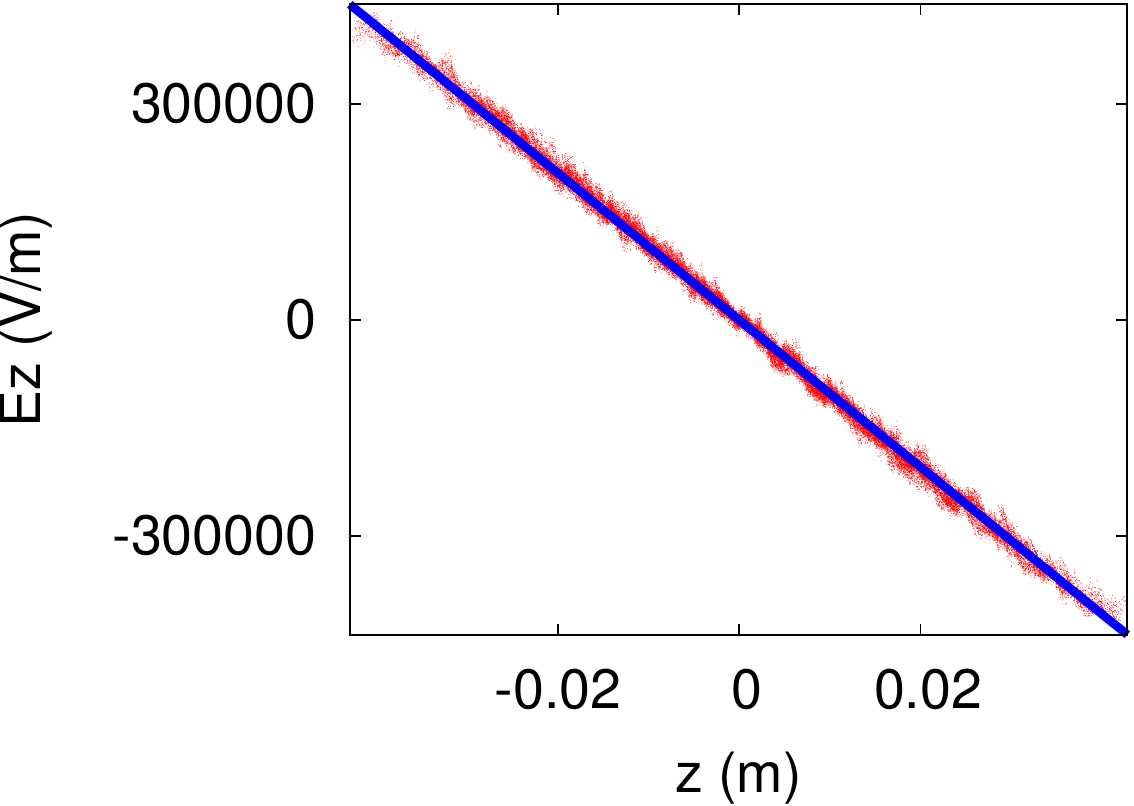}  
\caption{\label{fields}Transverse (top plot) and longitudinal (lower plot) electric fields experienced by the macropaticle simulated with {\sc elegant-bh} (symbols) and obtained from Eq.~\ref{fieldeqs} (lines). }
\end{figure}

We first  consider a 3D homogeneous ellipsoidal bunch with electric field linearly dependent on the position within the charge distribution as~\cite{Lapostolle:1965}
\begin{eqnarray}
\label{fieldeqs}
E_{u}(u) &=& \frac{C}{\gamma^2}\frac{(1-f)u}{r_{u}(r_{x}+r_{y})r_z}, \nonumber \\ 
E_z(z) &=& \frac{C f}{r_x r_y r_z}z, 
\end{eqnarray}
where $C\equiv 3Q/(4 \pi \epsilon_0)$, $u \in [x,y]$, $r_{x,y,z}$ are the ellipsoid semiaxes, $f \approx \sqrt{r_x r_y} / 3\gamma r_z$ and $Q$ is the bunch charge.

The simulated fields are in excellent agreement with the analytical expressions given by Eq.~\ref{fieldeqs} as shown on Fig.~\ref{fields}.

\begin{figure}[hhhh!!!!]
\includegraphics[width=0.98\linewidth]{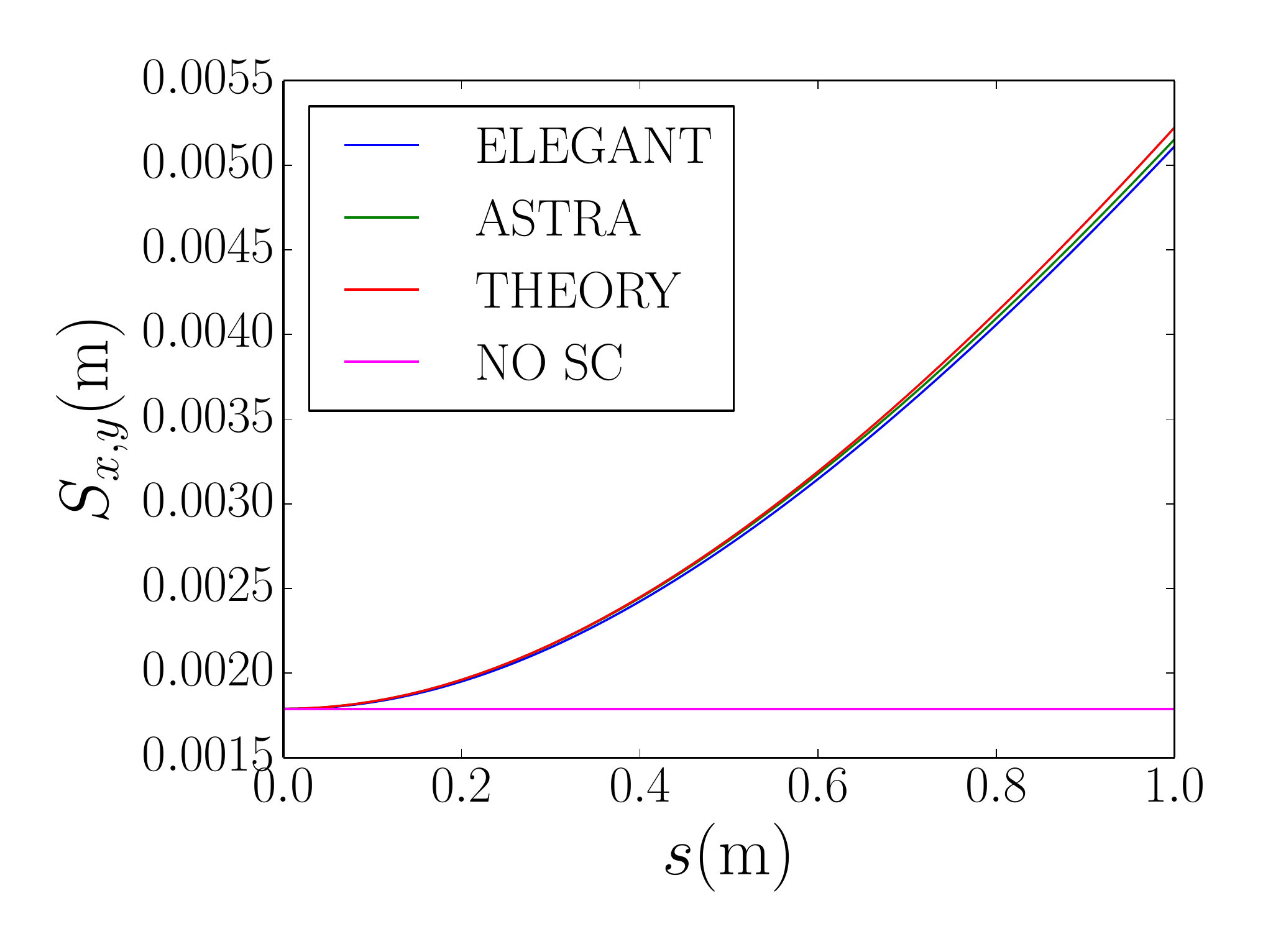}
\caption{\label{envelope} Comparison of the beam envelope evolution along a 1-m drift.  Eq.~\ref{K-V} solution (red), is compared against {\sc astra}  (green), and {\sc elegant-bh} 
simulation (blue). The magenta line corresponds to no space charge case.}
\end{figure}

To assess longer-term tracking, we compared the evolution of the 5 MeV, 500 A beam envelope over a drift space. 
For a stationary uniform beam the transverse envelope evolution is governed by~\cite{Kapchinsky:1959}
\begin{equation}
\label{K-V}
 S_{x,y}'' - \frac{\varepsilon^2_{rx,ry}}{S_{x,y}^3}-\frac{K_p}{2(S_{x,y}+S_{y,x})}=0,
\end{equation}
where $S_{x,y}$ is the rms beam size in ${x,y}$, $\varepsilon_{rx,ry}$ is the corresponding geometric 
emittance and $K$ is a space charge perveance defined (taken to be an order of $10^{-8}$), for a parabolic
current profile, as $K_p\equiv(QI\lambda)/(20\sqrt{5}\pi\epsilon_0mc^3\gamma^3\beta^2)$ where $Q$ is the bunch charge, $I$ the peak current, $\lambda$ the RF wavelength, and $m$ the electron mass.

Figure \ref{envelope} compares the solution of Eq.~\ref{K-V} against the beam envelope simulated with {\sc elegant-bh}. The geometric emittance was taken to be very low ($\varepsilon_{x,y} =1\times10^{-11}$~m) for these studies so that when space charge was turned off the beam envelope (dashed line) stayed quasi-constant. 

\begin{figure}[hhhhh!!!!]
\includegraphics[width=0.98\linewidth]{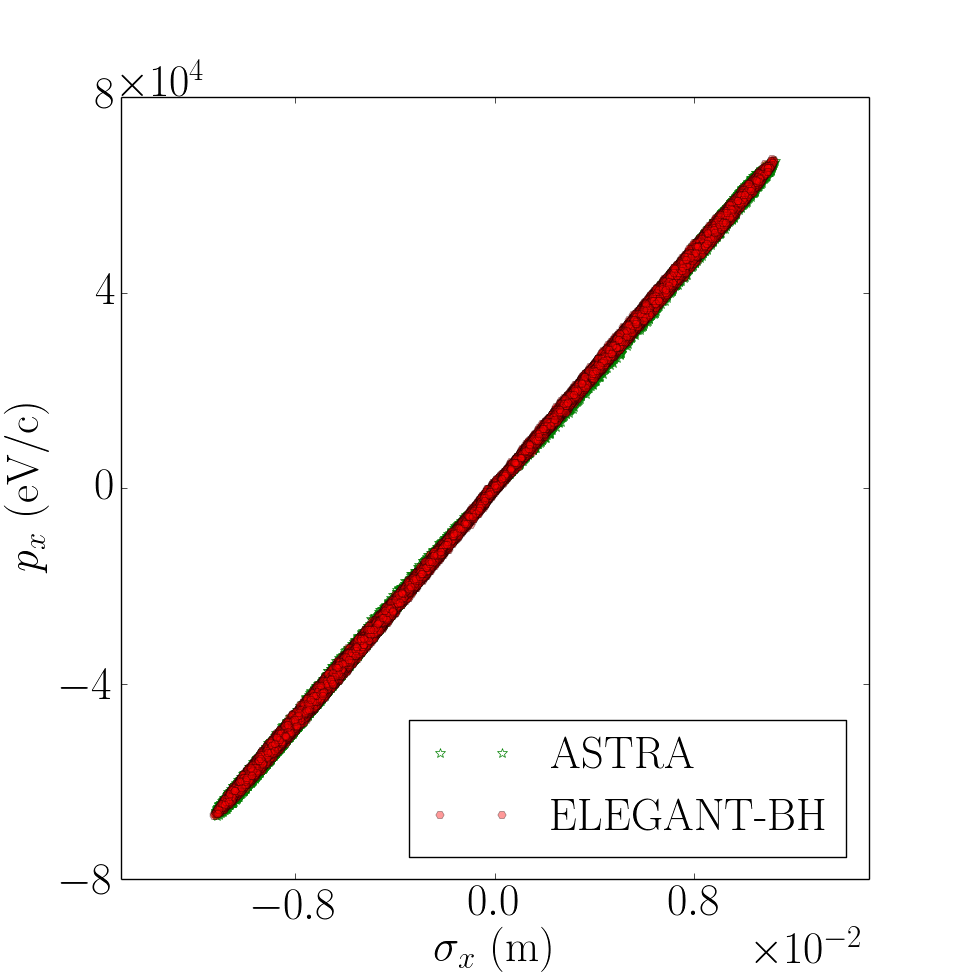}
\caption{\label{elastra} Comparison of the beam phase space in {\sc astra}  and {\sc elegant-bh} 
simulations.}
\end{figure}

Finally Fig.~\ref{elastra} compares the final transverse phase space at $s=1$~m obtained with {\sc elegant-bh} and {\sc astra} for the 50 MeV, 500 A beam. 
These two computer programs despite their very different space charge algorithms are in very close agreement. The code {\sc astra} relies on the same quasi-electrostatic approximation as used in {\sc elegant-bh} but implements a cylindrical-symmetric space charge algorithm using a $r-\phi$ two-dimensional grid for depositing the charge and solving Poisson's equation in the bunch frame~\cite{ASTRAmanual}.

%\bibliography{biblio_short1}
%\bibliographystyle{unsrt}

\end{document}